\documentclass[aps,pra,numerical,onecolumn,preprint,showpacs]{revtex4-1}
\usepackage[dvips]{graphicx}%
\usepackage{bm}%
\usepackage{amsmath}
\usepackage{dcolumn}
\usepackage{longtable}
\usepackage[normalem]{ulem} 
\usepackage[colorlinks=true,linkcolor=blue,citecolor=blue ,urlcolor=blue]{hyperref}

\draft

\begin{document}

\title{Extremal Density Matrices for Qudit States}

\author{Armando Figueroa}
\email{armando.figueroa@nucleares.unam.mx}
\author{Julio A.  L\'opez-Sald\'ivar} 
\email{julio.lopez@nucleares.unam.mx}
\author{Octavio Casta\~nos}
\email{ocasta@nucleares.unam.mx}
\author{Ram\'on L\'opez--Pe\~na}
\email{lopez@nucleares.unam.mx}
\affiliation{%
Instituto de Ciencias Nucleares, Universidad Nacional Aut\'onoma de M\'exico, Apartado Postal 70-543, 04510 M\'exico DF,   Mexico}

\date{\today}

\begin{abstract}
An algebraic procedure to find extremal density matrices for any Hamiltonian of a qudit system is established. The extremal density matrices for pure states provide a complete description of the system, that is, the energy spectra of the Hamiltonian and their corresponding projectors. For extremal density matrices representing mixed states, one gets mean values of the energy in between the maximum and minimum energies associated to the pure case. These extremal densities give also the corresponding mixture of eigenstates that yields the corresponding mean value of the energy. We enhance that the method can be extended to any hermitian operator. 

\end{abstract}

\maketitle


\section{Introduction}

The density matrix approach was introduced to describe statistical concepts in quantum mechanics by Landau~\cite{landau}, Dirac~\cite{dirac}, and von Neumann~\cite{vonneumann}.  In several branches of physics like polarized spin assemblies or qudit systems, and cavity electrodynamics the density matrix approach can be cast into a su(d) description~\cite{mahler}.  The Bloch vector parametrization was used to describe the 2-level problem which later on was generalized to describe beams of particles with spin $s$ in terms of what are known as Fano statistical tensors~\cite{fano,blum}. In particular $(2s+1)^2$ projectors defining the generators of a unitary algebra have been introduced in \cite{newton} to expand a density matrix of spin systems, even more, they established a procedure to reconstruct the density matrix by a finite number of magnetic dipole measurements with Stern-Gerlach analyzers and concluded that it was necessary to do at least $4s$ measurements to reconstruct the density matrix of pure states while $4s(s+1)$ were required for mixed states~\cite{newton,park}. An experimental reconstruction of a cavity state for $s=4$ using this method is given in~\cite{walser}. Another approach uses the Moore - Penrose pseudoinverse to express the elements of the spin density matrix in terms of $(2s+1)(4s+1)$ probabilities of spin projections~\cite{klose}. A method to reconstruct any pure state of spin in terms of coherent states is provided in~\cite{amiet} and by means of non orthogonal projectors on coherent states a reconstruction of mixed states can be done~\cite{amiet2}. A parametrization based on Cholesky factorization~\cite{horn} was first used to guarantee the positivity of the spin density matrices in~\cite{chung}, and more recently, a tomographic approach to reconstruct 
them~\cite{dodonov,olga,lopez1,lopez2}. 

In the last twenty years, a lot of work related with parametrization of the density matrices of $d$-level quantum systems has been done~\cite{bertlmann,hioe,aktha,bruning,jarls}. This is due to its applications to quantum computation and quantum information systems~\cite{nielsen}. The decomposition of the density matrix into a symmetrized polynomial in Lie algebra generators has been determined in~\cite{ritter}. A novel tensorial representation for density matrices of spin states, based on  Weinberg's covariant matrices, may be another important generalization of the Bloch sphere representation~\cite{giraud}. 

Actually, there are several parametrizations of finite density matrices:  generalizations of the Bloch vector~\cite{bertlmann}, the canonical coset decomposition of unitary matrices~\cite{aktha,bruning}, the recursive procedures to describe $n \times n$ unitary matrices in terms of those of $U(n-1)$\cite{jarls,fujii}, by factorizing $n \times n$ unitary matrices in terms of points on complex spheres~\cite{dita}, and by defining generalized Euler angles~\cite{tilma}. Even in the case of composite systems there are parametrizations of finite density matrices~\cite{spengler,petru}.
 
Recently we have established a procedure to determine the extremal density matrices of a qudit system associated to the expectation value of any observable~\cite{figueroa}. These matrices provide an extremal description of the mean values of the energy, and in the case of restricting them to pure states the energy spectrum is recovered. So, apart from being an alternative tool to find the eigensystem one has information of mixed states which minimize its mean value.

The aim of this work is to give another option to compute extremal density matrices in a qudit space by means of an algebraic approach that leads to an underdetermined linear system in terms of the components of the Bloch vector $\boldsymbol{\lambda}=(\lambda_1, \lambda_2, \ldots, \lambda_{d^2-1})$, the antisymmetric structure constants $f_{ijk}$ of a $su(d)$ algebra, and the parameters of the Hamiltonian operator $ \{ h_k \}$. Their solution, in general, implies to get the Bloch vector in terms of a known number of free components. These are determined by establishing a system of equations associated to the characteristic polynomial of the density matrix. Finally, one arrives to the extremal density matrices of the expectation value of the Hamiltonian, which for the pure case let us obtain the corresponding full spectrum or for the mixed case at most $ d\, !$ extremal mean value energies. Another goal is to bring and join different algebraic tools in the study of the behaviour of both the density matrix and hermitian operators.

\section{Generalized Bloch-vector parametrization}
Any hermitian Hilbert-Schmidt operator acting on the $d$-dimensional Hilbert space can be expressed in terms of the identity operator plus a set of hermitian traceless operators $\{\hat{\lambda}_1\ldots\hat{\lambda}_{d^2-1}\}$ which are the generators of the $su(d)$ algebra. In this basis, the Hamiltonian operator $\hat{H}$ and the density matrix $\hat{\rho}$ are written as~\cite{kimura}
\begin{eqnarray}
\hat{H} &=& \frac{1}{d} h_0\widehat{I}+ \frac{1}{2} \sum_{k=1}^{d^2-1} h_k\, \hat{\lambda}_k \, , \label{s1:eq1}\\
\hat{\rho} &=& \frac{1}{d}\hat{I}+ \frac{1}{2}\sum_{k=1}^{d^2-1}\lambda_k \, \hat{\lambda}_k \, , \label{s1:eq2}
\end{eqnarray}
with the definitions $h_0 \equiv {\rm Tr}(\hat{H}) , \,h_k \equiv {\rm Tr} 
(\hat{H}\hat{\lambda}_k)$ and $\lambda_k \equiv {\rm Tr} (\hat{\rho}\hat{\lambda}_k)$. 

These generators are completely characterized by means of their commutation and anticommutation relations given by
\begin{eqnarray}
\left[\hat{\lambda}_{j},\hat{\lambda}_{k}\right] &= & 2 \, i \sum_{q=1}^{d^2-1} f_{jkq} \hat{\lambda}_{q} \, , \label{s1:eq3}\\
\{\hat{\lambda}_{j},\hat{\lambda}_{k}\} &= &\frac{4}{d}\delta_{jk}\hat{I} + 2 \sum_{q=1}^{d^2-1} d_{jkq} \hat{\lambda}_{q} \, , \label{s1:eq4}
\end{eqnarray}
where $d_{jkq}$ and $f_{jkq}$ are the symmetric and antisymmetric structure constants
\begin{eqnarray}
d_{jkq} & = & \frac{1}{4} {\rm Tr} (\{\hat{\lambda_j} , \hat{\lambda_k} \} \hat{\lambda_q}) \, , \label{s1:eq5} \\
f_{jkq} & = & \frac{1}{4 i} {\rm Tr} (\left[ \hat{\lambda_j} , \hat{\lambda_k} \right] \hat{\lambda_q}) \, , \label{s1:eq6}
\end{eqnarray}
and consequently, it follows the multiplication law~\cite{macfarlane}
\begin{eqnarray}
\hat{\lambda}_{j} \, \hat{\lambda}_{k} = \frac{2}{d} \hat{I} \, \delta_{jk}
+ \sum_{q=1}^{d^2-1} \left( d_{jkq} + i  f_{jkq}\right) \hat{\lambda}_{q} \, .  \label{s1:eq6a}
\end{eqnarray}

A realization of the generators can be given by the generalized Gell-Mann matrices~\cite{hioe}, consisting in $s= 1,\ldots,\frac{d(d-1)}{2}$ symmetric matrices
\begin{equation}\label{s1:eq7}
\hat{\lambda}_{s}=\hat{P}_{jk}+ \hat{P}_{kj}\, , 
\end{equation}
plus $a= \frac{d(d-1)}{2}+1,\ldots,d(d-1)$ antisymmetric matrices 
\begin{equation}\label{s1:eq8}
\hat{\lambda}_{a}=- i(\hat{P}_{jk} - \hat{P}_{kj} ) \, , 
\end{equation}
and $l=1,\ldots,d-1$ diagonal ones
\begin{eqnarray}\label{s1:eq9}
\hat{\lambda}_{d(d-1)+l}&=&\sqrt{\frac{2}{l(l+1)}}(\hat{P}_{11} + \hat{P}_{22} + \cdots + 
 \hat{P}_{l \, l} - l\hat{P}_{l+1\, l+1}) \, ,
\end{eqnarray}
where $1 \leq j < k \leq d$ and $\hat{P}_{jk}\equiv|j\rangle \langle k|$ are matrices with $1$ in the component $(j,\, k)$ and $0$ otherwise.

This type of realization belongs to the so called generalized Bloch vector parametrization~\cite{bertlmann}. The Fano statistical tensors~\cite{fano,blum}, the multipole moments~\cite{newton}, the Weyl matrices~\cite{bertlmann}, and the generalized Gell-Mann matrices~\cite{hioe}, belong to this group. Therefore a vector with $d^2-1$ real components define the so called generalized Bloch vector~\cite{mahler,hioe},
\begin{eqnarray}\label{s1:eq10}
\boldsymbol{\lambda}&=&(\lambda_1, \ldots, \lambda_{\frac{d(d-1)}{2}}, \lambda_{\frac{d(d-1)}{2}+1}, \ldots, 
\lambda_{d(d-1)}, \lambda_{d(d-1)+1}, \ldots, \lambda_{d^2-1}) \, ,
\end{eqnarray}
whose magnitude is bounded by~\cite{kimura2}
\begin{equation}\label{s1:eq11}
 | \boldsymbol{\lambda} | \leq \sqrt{\frac{2(d-1)}{d}} \, ,
\end{equation}
where the equality specifies a necessary condition to represent a pure state.

In general, a $SU(d)$ unitary transformation acting on a hermitian matrix implies a rotation in its components, i.e.,
\begin{eqnarray}\label{s1:eq12}
\hat{H}^{\prime}  &=&  \hat{U}  \, \hat{H} \, \hat{U}^{\dagger} =  \frac{1}{d} h_0 \hat{I}+ \frac{1}{2}\sum_{j=1}^{d^2-1} h_j \, \hat{U} \, \hat{\lambda}_j  \, \hat{U}^{\dagger} \nonumber \\
&\equiv&  \frac{1}{d} h_0 \hat{I}+ \frac{1}{2}\sum_{j=1}^{d^2-1} h_j^{\prime} \,  \hat{\lambda}_j   \, ,
\end{eqnarray}
where in the last equality, one has defined
\begin{eqnarray}
h_k^{\prime} &=& {\rm Tr}(\hat{H}^{\prime} \, \hat{\lambda}_k ) 
=  \sum_{j=1}^{d^2-1} O_{kj}\, h_j \, , \label{s1:eq13}
\end{eqnarray}
and 
\begin{equation}\label{s1:eq14}
O_{kj} \equiv \frac{1}{2} {\rm Tr}(\hat{\lambda}_k \hat{U} \, \hat{\lambda}_j  \, \hat{U}^{\dagger} )  \, ,
\end{equation}
are elements of an orthogonal matrix that belongs to the $SO(d^2-1)$ group, which provides the adjoint representation of $SU(d)$~\cite{mallesh,bengtsson}.

\section{Positivity conditions for the Density Operator}

The density matrix must satisfy the following three properties: (a) It is Hermitian, (b) it has trace one, and (c) all its eigenvalues are positive semidefinite. While for dimension $d=2$, the condition ${\rm Tr}(\hat{\rho}^2) \leq 1 $ implies (c), for $d \geq 3$ that is not true.

The positivity conditions of the density matrix are established by the set $\{ a_k \}$ of coefficients of its corresponding characteristic polynomial. This set can be obtained by means of the recursive relation known as Newton-Girard formulas~\cite{bruning,seroul}
\begin{equation}\label{s2:eq2}
a_k = \frac{1}{k}  \sum^{k}_{j=1} (-1)^{j-1} \, a_{k-j} \, t_j \, ,
\end{equation}
with the definitions $a_0 = a_1 = 1$, $a_d = \det \hat{\rho}$, and $ t_j = {\rm Tr} (\hat{\rho}^j )$, for $j=1,\ldots , d $. Therefore the allowed density matrix must satisfy the following system of $d-1$ simultaneous polynomial equations
\begin{equation}\label{s2:eq3}
a_k = c_k \, , \qquad {\rm for} \quad k=2,3,\ldots,d \, ,
\end{equation}
where the constants $c_k$ fix the degree of mixing of the system. Thus, they must be in the region given by~\cite{kimura,byrd,deen}
\begin{equation}\label{s2:eq1}
0 \leq c_k \leq \frac{1}{d^k} {d \choose k} \, ,
\end{equation}  
where ${d \choose k}$ denotes a binomial coefficient. The upper bound defines the most mixed state and then it has maximum entropy, while the lower bound specifies pure states  which have zero entropy. Additionally, the $ c_k =0 $ for $k > {\rm rank }( \hat{\rho})$~\cite{horn}.

All of them are polynomial functions in terms of the invariants of the density matrix, i.e.,  $ t_j $, for $j=1,\ldots , d $. In terms of $ t_k \equiv {\rm Tr} (\hat{\rho}^k ) $, it is defined the symmetric matrix called Bezoutian~\cite{schwarz,procesi,gerdt} 
\begin{equation} 
\boldsymbol{B}_d = \left(
\begin{array}{ccccc}
	d      & t_1  	& t_2 		& \cdots 	& t_{d-1} 	\\
	t_1    & t_2  	& t_3 		&  \ddots 	& t_d   	\\
	t_2    & t_3   	& \ddots   	&        	& t_{d+1}   \\
	\vdots & \ddots &     		&        	& \vdots     \\
	t_{d-1} & t_d 	& t_{d+1} 	& \cdots 	& t_{2(d-1)}
\end{array} \right) \, . \label{B:eq1}
\end{equation}
A polynomial with real coefficients has reals roots iff the Bezoutian matrix is positive definite~\cite{procesi}. Hence, the compatible region among the invariants is obtained with the intersection of the positivity conditions of the density matrix from~(\ref{s2:eq1}) with the respective positivity conditions of the Bezoutian (see details in Appendix~\ref{B}).

\section{Rayleigh Quotient and the Density Matrix} \label{Ray}

The Rayleigh quotient $\mathbf{R}_T(\psi)$ of a hermitian Hilbert-Schmidt operator $\hat{T}$ is 
\begin{eqnarray}
\mathbf{R}_T(\psi) := \frac{\langle \psi|  \hat{T}  |\psi \rangle}{\langle \psi|  \psi \rangle} \, , \label{s3:eq1}
\end{eqnarray}
where $|\psi \rangle$ is a $d$-dimensional complex vector. Since the Rayleigh quotient is invariant under scale transformations, in searching its maximum or minimum it suffices to confine the search on unit norm vectors, i.e., when $ \langle \psi|  \psi \rangle  = 1$~\cite{lax}. This leads to define the numerical range $W(\hat{T})$, which is the set of all possible Rayleigh quotients $ \mathbf{R}_T(\psi)$ over the unit vectors:
\begin{equation}\label{s3:eq2}
W(\hat{T}) = \{ \mathbf{R}_T(\psi); \quad  \langle \psi|  \psi \rangle  = 1 \} \, .
\end{equation}

The numerical range $W(\hat{T})$ is a closed interval on the real axis, whose end points are the extreme eigenvalues of $\hat{T}$~\cite{bhatia}. This result is a particular case of the Courant-Fischer Theorem~\cite{horn}, which states that every eigenpair (eigenvalue and eigenvector) of $\hat{T}$ is the solution of a optimization (max-min problem) of $W(\hat{T})$ in some subspace of $\hat{T}$. Therefore, eigenvectors and eigenvalues of $\hat{T}$ are the critical points and critical values, respectively, of the Rayleigh quotient and $W(\hat{T})$ is the convex hull of its eigenvalues. 

In the density matrix formalism, the numerical range of the Hamiltonian (or any hermitian operator) can be identified with its mean value in an arbitrary state $ \hat{\rho} $, i.e., $ \langle \hat{H} \rangle = {\rm Tr}( \hat{H}\, \hat{\rho} ) $~\cite{gawron}. In this scheme, a useful theorem is the following one. 

{\bf Theorem 1}~\cite{horn}. {\it Let $\hat{H} $ and $ \hat{\rho} $ be $d \times d$ hermitian matrices with their eigenvalues $\{ \epsilon_j \}$ and $ \{ \gamma_k \} $, respectively, arranged in descending order, v.g., $\epsilon_1 \geq \epsilon_2 \cdots \geq \epsilon_d$ and $\gamma_1 \geq \gamma_2 \cdots \geq \gamma_d$. Thus, one has the inequality
\begin{equation} \label{s3:eq16a}
\sum_{i=1}^d \epsilon_{d-i+1} \, \gamma_{i} \leq {\rm Tr} (\hat{H} \, \hat{\rho}) \leq \sum_{i=1}^d \epsilon_{i} \, \gamma_{i} \, .
\end{equation}
If either inequality is an equality, then $\hat{H}$ and $\hat{\rho}$ commute. }  

Since the equality is easy to verify when $\hat{H}$ and $\hat{\rho}$ are diagonals (diagonal frame), this theorem leads to the assumption that the density matrix can be adapted to get the spectrum of $\hat{H}$ if they both commute. In that way, a related fact is the following. 

{\bf Proposition 1.} {\it For an arbitrary $\hat{\rho}^c$ commuting with $\hat{H}$, in any frame $ \langle \hat{H} \rangle^c $ depends on at most $d-1$ variables parametrizing $\hat{\rho}^c$.}

{\it Proof}. Since $\hat{\rho}^c$ and $\hat{H}$ commute, they are simultaneously diagonalizable therefore, in the diagonal frame, 
\begin{eqnarray}
\langle \hat{H} \rangle^c = \frac{1}{d} h_0 + \frac{1}{2} \boldsymbol{h}^{\prime} \cdot \boldsymbol{\lambda}^{\prime} \, ,\label{s3:eq17t1}
\end{eqnarray}
where we use the convention
\begin{eqnarray}
 \boldsymbol{h}^{\prime} = (0, \ldots, 0, h^{\prime}_{d(d-1)+1}, \ldots, h^{\prime}_{d^2-1}) \, , \nonumber \\
 \boldsymbol{\lambda}^{\prime} = (0, \ldots, 0, \lambda^{\prime}_{d(d-1)+1}, \ldots, \lambda^{\prime}_{d^2-1}) \, , \nonumber
\end{eqnarray}
and $h_0 = {\rm Tr}(\hat{H})$. By applying the relation~(\ref{s1:eq13}) one has $\boldsymbol{\lambda}= \boldsymbol{O} \boldsymbol{\lambda}^{\prime}$ and $\boldsymbol{h}= \boldsymbol{O} \boldsymbol{h}^{\prime}$ with $ \boldsymbol{O} $ as the orthogonal matrix from~(\ref{s1:eq14}). Setting aside scalar matrices, since orthogonal transformations preserve the dot product, $\langle \hat{H} \rangle^c$ is an invariant quantity and depends on at most $d-1$ variables of $\hat{\rho}^c$. {\it q.e.d.}

With the aim to provide further properties of the set of density matrices that commute with $ \hat{H}$ and propose an algorithm to compute the extremal mean values of the energy in the density matrix formalism, we consider from here on $\hat{H}$ as a given non-scalar matrix and establish the proposition:

{\bf Proposition 2.} {\it Let $\hat{H}$ and $ \hat{\rho} $ be two finite $d \times d $ hermitian matrices where $ \hat{\rho} $ represents an arbitrary density matrix. For a fixed degree of mixture, the critical points of $ \langle \hat{H} \rangle $ determine the extremal density matrices $\hat{\rho}^c_m$ commuting with $\hat{H} $ and $  \langle \hat{H} \rangle^c = {\rm Tr} ( \hat{H} \,  \hat{\rho}^c_m ) $, with $ 1 \leq m \leq d\,!$. }

{\it Proof}. Suppose that $\hat{\rho}$ is unitarily related to a density matrix $\hat{\rho}^c$ which commute with $\hat{H}$ then, $\hat{\rho} = \hat{U}(\theta_m)  \, \hat{\rho}^c \, \hat{U}^{\dagger}(\theta_m) $, where  $\hat{U}(\theta_m)$ define a $SU(d)$ unitary transformation. Therefore, with the mean value $ \langle \hat{H} \rangle = {\rm Tr} ( \hat{H} \,  \hat{\rho}) $ one can define the scalar function
\begin{equation}\label{s3:eq8}
 E( \theta_m , \lambda_k^c , h_i) \equiv {\rm Tr}( \hat{H}\, \hat{U}(\theta_m)  \, \hat{\rho}^c \, \hat{U}^{\dagger}(\theta_m))   \, ,  
\end{equation}
where the real constants $ h_i $ and $\lambda_k^c $ are the components of the expansion of $\hat{H}$ and $ \hat{\rho}^c $ respectively, in a basis for the Hilbert space of Hermitian operators. 

Otherwise, if $\hat{U}$ is sufficiently close to the identity, by considering $ \{\theta_m\}$ as infinitesimal parameters one can make a Taylor series expansion of the function as follows
\begin{eqnarray}
E ( \theta_m , \lambda_k^c , h_i) &=& E ( 0 ,\lambda_k^c , h_i) + \sum_{p=1}^{d^2-1} \theta_p \, \epsilon_p  
+ \frac{1}{2} \sum_{q,p=1}^{d^2-1} \theta_p \theta_q \, \epsilon_{p,q} + \mathcal{O}( \theta^3 ) \, , \label{s3:eq9} 
\end{eqnarray}
where we have defined
\begin{eqnarray}
\epsilon_q & \equiv & \frac{\partial }{\partial \theta_q} E ( \theta_m , \lambda_k^c , h_i) \bigg|_{ \{\theta_m\} \to 0} \, , \label{s3:eq10}  \\
\epsilon_{p,q} & \equiv & \frac{\partial^2 }{\partial \theta_p \theta_q} E ( \theta_m , \lambda_k^c , h_i) \bigg|_{ \{\theta_m\} \to 0} \, . \label{s3:eq11} 
\end{eqnarray}

As the $SU(d)$ unitary transformation is infinitesimal, one has that
\begin{eqnarray}
\hat{\rho} &\approx& \hat{\rho}^c + i \sum_{p=1}^{d^2-1} \theta_p [\hat{\lambda_p} \,,\, \hat{\rho}^c ] 
+\frac{i^2}{2} \sum_{q,p=1}^{d^2-1} \theta_q \theta_p [ \hat{\lambda}_q \,,\, [\hat{\lambda}_p  \,,\, \hat{\rho}^c ] ] \, . \label{s3:eq12} 
\end{eqnarray}
Substituting the last expression into~(\ref{s3:eq8}), comparing with~(\ref{s3:eq9}) and by applying the cyclic property of the trace, Eqs.~(\ref{s3:eq10}) and (\ref{s3:eq11}) lead to
\begin{eqnarray}
\epsilon_q & = & i \, {\rm Tr } \left( [ \hat{\rho}^c \,,\, \hat{H} ] \,  \hat{\lambda}_q \right) \, ,\label{s3:eq13} \\
\epsilon_{p,q} & = & i^2 {\rm Tr } \left( [\hat{\lambda}_p  \,,\, \hat{\rho}^c ]\, [ \hat{\lambda}_q \,,\, \hat{H}] \right) \, . \label{s3:eq14}
\end{eqnarray}

The algebraic system which determines the critical points is given by equating Eq.~(\ref{s3:eq13}) to zero, for $q=1,\ldots, d^2-1$, whereby $[ \hat{\rho}^c \,,\, \hat{H} ] = \boldsymbol{0}$. Hence, $\langle \hat{H} \rangle$ achieves its extreme values at $\hat{\rho}^c$. In that sense, any density matrix which commutes with $\hat{H}$ and optimizes its mean value, is extremal. Even though the commutativity is satisfied by hypothesis, it implies that any state $\hat{\rho}$ can be approximated at first-order by $ \hat{\rho}^c $ and $\langle \hat{H} \rangle$ has an error that vanishes to the second-order in $\mathcal{O}( \theta^2 ) $, i.e.,
\begin{eqnarray}
\langle \hat{H} \rangle \approx \langle \hat{H} \rangle^c + \mathcal{O}( \theta^2 ) \, .
\end{eqnarray}
Thus, by means of the proposition $1$, $ \hat{\rho}^c $ depends in general on $d-1$ variables that are fixed by establishing a degree of mixture through the expressions~(\ref{s2:eq3}).

Finally, the highest degree of the polynomial $a_k$ in~(\ref{s2:eq3}) is $k$. Then, by Bezout's theorem, the number of solutions for the polynomial system (known as Bezout bound or Bezout number) is at most the product of the degree of all the equations, i.e., $\Pi_{k=2}^d k = d\,!$~\cite{waerden,cox,gelfand,sturmfels}. All this implies that the single critical matrix $\hat{\rho}^c$ represents at most $d\,!$ different critical density matrices $\hat{\rho}^c_m$, where $1\leq m \leq d\,!$. {\it q.e.d.}

By matching results, in the non degenerate case of $ \hat{H} $, if all $\hat{\rho}^c_m$ are pure states, they must correspond to one-dimensional eigenprojectors of $ \hat{H} $.

\section{Algebraic approach to extremal density matrices}\label{extremal}

In a previous work~\cite{figueroa} we proposed an approach to obtain information of the energy spectrum of a Hamiltonian  by considering its mean value together with $d-1$ constraints to guarantee the positivity of the density matrix. This is achieved by defining the function
\begin{equation}\label{s3:eq5}
f(\lambda_k,  \Lambda_j, h_i, c_l ) \equiv {\rm Tr}(\hat{H} \, \hat{\rho}) + \sum_{j=2}^d \Lambda_j (a_j - c_j ) \, ,
\end{equation}
which depends on $d^2$ real parameters $\{ h_i \}$ and $d^2-1$ independent variables $\{ \lambda_k \} $ associated to the expansions~(\ref{s1:eq1}) and~(\ref{s1:eq2}), respectively. Additionally, there are $d-1$ Lagrange multipliers $\{ \Lambda_j \}$ and $d-1$ positive real constants $\{c_j\}$ to fix the degree of purity of the density matrix (see the bound~(\ref{s2:eq1})). One can note that $f(\lambda_k,  \Lambda_j, h_i, c_l )$ is a continuous function because is the sum of the Rayleigh quotient $\mathbf{R}_H(\psi)$, where $\hat{\rho} \equiv |\psi \rangle  \langle \psi |$ with $ {\rm Tr} \hat{\rho} = 1 $, and the positivity constraints which are polynomials in the variables of the density matrix. Therefore, in order to reach all the eigenvalues of $ \hat{H}$ and its numerical range, one must find the min-max sets of $f(\lambda_k,  \Lambda_j, h_i, c_l )$ with respect to the variables $\{ \lambda_k \} $ and the Lagrange multipliers $\{ \Lambda_j \}$. Their respective derivatives give $d^2+d-2$ algebraic equations, 
\begin{eqnarray}
\frac{1}{2} h_q  - \sum^d_{j=2} \Lambda_j \frac{\partial a_j}{\partial \lambda_q} = 0  \, , & & \quad  q=1, \dots d^2-1 \, , \label{s3:eq6}\\
a_p = c_p \, , & & \quad  \, p=2, \dots d \, . \label{s3:eq7}
\end{eqnarray}
These sets of algebraic equations determine the extremal
values of the density matrix, i.e., $\lambda_q=\lambda_q^c$ and $\Lambda_q=\Lambda_q^c$ for which the expressions~(\ref{s3:eq6}) and~(\ref{s3:eq7}) are satisfied. By substituting $\lambda_q^c$ into equation~(\ref{s1:eq2}) one obtains the extremal density matrices. If we restrict the solutions to pure states $\{ c_p = 0 \}$, we have shown explicitly that the energy spectrum of the Hamiltonian is recovered for $d = 2$ and $3$~\cite{figueroa}. Extremal expressions for the mean value of the Hamiltonian can be obtained with density matrices representing mixed quantum states, which determine also the corresponding mixture of eigenstates of the Hamiltonian.

From here on, we describe an alternative algebraic procedure to get the extremal density matrices which is simpler than the one mentioned above. First of all, notice that propositions $1$ and $2$ in section~\ref{Ray} are based on the assumption of a common basis, which it is always possible to find if $ \hat{\rho}^c$ and $\hat{H}$ commute. Thus, in the following paragraphs, we propose for the pure case (or mixed case) a systematic approach to get information about the Hamiltonian spectrum (or mean value of the Hamiltonian), i.e., its numerical range (interval of extremal mean values of $\hat{H}$), without making use of a diagonalization procedure.  

First we replace into the commutator $[ \hat{H}, \hat{\rho}]=0$ the expressions Eqs.~(\ref{s1:eq1}) and~(\ref{s1:eq2}) and use the properties of the generators $\hat{\lambda}_q $ of the $su(d)$ algebra. Then the expression~(\ref{s3:eq13}) gives rise to the $d^2-1$ dimensional homogeneous system of equations 
\begin{eqnarray}
\boldsymbol{M} \cdot \boldsymbol{\lambda} &=& 0  \, , \label{s3:eq15}
\end{eqnarray}
that determines the critical points and where $\boldsymbol{\lambda} $ is the Bloch vector defined in~(\ref{s1:eq10}). The matrix elements of the skew symmetric matrix $\boldsymbol{M}$ of $d^2-1$  dimensions are given by 
\begin{eqnarray}
\boldsymbol{M}_{i,j}= \sum_{k=1}^{d^2-1} f_{i \, j \,k} \, h_k \, , \label{s3:eq16}
\end{eqnarray}
where $f_{i \, j \, k}$ are the antisymmetric structure constants of the $su(d)$ algebra. 

A single solution of the homogeneous system~(\ref{s3:eq15}) can be obtained through the Gauss-Jordan elimination method and it is identified as the critical Bloch vector $\boldsymbol{\lambda}^c$, with its $ n $ free variables equal to the dimension of the null space of $ \boldsymbol{ M} $. This implies that maximal mixed states are always critical for any observable because the null space always contains the zero vector. 

On the other hand, notice that $\widehat{w}_q \equiv i\,[ \widehat{H} \,,\, \hat{\lambda}_q ] $ are hermitian vectors spanning the tangent space of the orbits associated with $\hat{H}$, with $q=1,2,\ldots,d^2-1$. Then by substituting the Hamiltonian expression~(\ref{s1:eq1}) one gets
\begin{equation}
\hat{w}_q = \sum_{k,l} f_{qkl}\, h_k \, \hat{\lambda}_l \, .
\end{equation}
Note that these vectors give the rows of $ \boldsymbol{ M} $, and the number of independent vectors $r$ is determined by the rank of the Gram matrix 
\begin{equation}
G_{q,p} = {\rm Tr} (\hat{w}_q \, \hat{w}_p )= 4 \, \sum_{k_1,k_2, j} f_{q k_1j} \, f_{p k_2 j} \, h_{k_1} \, h_{k_2} \, .
\end{equation}
Therefore $r$ determines the dimension of the tangent space of the Hamiltonian orbits and the rank of $ \boldsymbol{ M} $, i.e., $ r = {\rm rank} (\boldsymbol{ M} )$~\cite{helgason,kus,linden1,ercolessi,schirmer,gantmacher}. Furthermore, the maximal dimension of the orbit occurs when the Hamiltonian is non degenerate, i.e., when $r=d(d-1)$ and by comparing $d^2-1$ with $r$ one has that the system~(\ref{s3:eq15}) is always underdetermined with $n=d^2-1-r$ free variables (see Table~\ref{tab}).
\begin{table*}[h!]
\caption{ \label{tab}
Manifolds and their dimension for the unitary orbits of the Hamiltonian based on their diagonal form. It is supposed that $ \alpha > \beta > \gamma > \delta $. The number of free parameters of Eq.~(\ref{s3:eq15}) is only determined by the expression $n=d^2-1-r$~\cite{schirmer}.}
\medskip
\resizebox{0.75 \textwidth}{!} {
\begin{tabular}{ c | c  c  c  }
\hline
\hline
\textbf{Hamiltonian} & \textbf{Diagonal} &  \textbf{Manifold} & \textbf{Manifold} \\
\textbf{dimension} & \textbf{representation} &  & \textbf{dimension} \\
$ \boldsymbol{ d} $ &  &  & $ \boldsymbol{ r} $ \\
\hline
\hline
\hline
    	    & ${\rm diag} (\alpha , \alpha )$ & point & \textbf{0}  \\
            \textbf{2} & $ {\rm diag} (\alpha , \beta )$ & $U(2)/[U(1)\times U(1)]$ & \textbf{2} \\
\hline
    	    & ${\rm diag} (\alpha , \alpha , \alpha )$ & point & \textbf{0}  \\
\textbf{3} & $ {\rm diag} (\alpha , \beta , \beta )$ & $U(3)/[U(1) \times U(2)] $ & \textbf{4} \\
    		& $ {\rm diag} (\alpha , \beta , \gamma )$ & $U(3)/[U(1) \times U(1) \times U(1) ] $ & \textbf{6} \\
\hline
    		& ${\rm diag} (\alpha , \alpha , \alpha, \alpha )$ & point & \textbf{0}  \\
   		& $ {\rm diag} (\alpha , \beta , \beta, \beta )$ & $U(4)/[U(1) \times U(3)] $ & \textbf{6} \\
\textbf{4}  & $ {\rm diag} (\alpha , \alpha , \beta, \beta )$ & $U(4)/[U(2) \times U(2) ] $ & \textbf{8} \\
    		& $ {\rm diag} (\alpha , \beta , \gamma , \gamma )$ & $U(4)/[U(1) \times U(1) \times U(2) ] $ & \textbf{10} \\
    		& $ {\rm diag} (\alpha , \beta , \gamma , \delta )$ & $U(4)/[U(1) \times U(1) \times U(1) \times U(1) ] $ & \textbf{12} \\
\hline
\end{tabular}}
\end{table*}

To clarify the method, we are going to discuss the non degenerate and degenerate cases of $\hat{H}$ separately. In section~\ref{examples} we shall illustrate the method for quantum systems of dimensions $d=2, \, 3$ and $4$. 

\section{Non degenerate case of $\hat{H} $}

In this case, the rank of the matrix $ \boldsymbol{M} $ is given by $ r = d(d-1)$ and the $n$ free variables reach its minimum number, i.e., $ n = d-1 $. Therefore, $\boldsymbol{\lambda}^c$ is given by
\begin{equation}\label{s3:eq17}
\boldsymbol{\lambda}^c = (\lambda_1^c,\, \ldots,  \lambda_{d(d-1)}^c,\, \lambda_{d(d-1)+1},\ldots \lambda_{d^2-1} ) \, ,
\end{equation} 
where the $ d(d-1)$ components $\{ \lambda_q^c \}$ are functions of the parameters of the Hamiltonian, the antisymmetric structure constants and a set of $d-1$ independent free variables. It is natural to choose this set from the diagonal generators~(\ref{s1:eq9}) of $su(d)$.

The substitution of $\boldsymbol{\lambda}^c$ in~(\ref{s1:eq2}) gives its associated critical density matrix denoted as $\hat{\rho}^c$. Therefore, the determination of the $d-1$ free variables is done by solving the system of $d-1$ polynomial equations~(\ref{s2:eq3}), which by proposition 2 has at most $d\,!$ different solutions. This is also in agreement with the analysis in the diagonal representation $\hat{\rho}^c_{diag}$, wherein the action of the permutation group of $n$ elements produces $d \, !$ matrices~\cite{bengtsson,boya}. They satisfy the same polynomial system~(\ref{s2:eq3}) but give different mean values of $\hat{H}$. Therefore, the critical density matrices are given by
\begin{eqnarray}
\hat{\rho}^c_m &=& \frac{1}{d}\hat{I}+ \frac{1}{2}\sum_{k=1}^{d^2-1}\lambda_{m,\,k}^c \, \hat{\lambda}_k \, ,  \label{s3:eq18}
\end{eqnarray}
with $m$ denoting the Bloch vector solution, the variables $\{\lambda_{m,\,k}^c\}$ are function only of the known quantities, i.e., the structure constants, the parameters of the Hamiltonian $(h_0, h_k)$, and $d-1$ constants $\{ c_k \}$. The number $m$ of solutions decreases up to $d$ when~(\ref{s3:eq18}) represent pure states (all $ \{ c_k = 0\}$) and the extremal density matrices are one-dimensional orthogonal projectors. 

As to be expected, the expectation value of the Hamiltonian is in general given by
\begin{equation} \label{s3:eq19}
\langle \hat{H} \rangle_m^c \equiv {\rm Tr}( \hat{H}\,  \hat{\rho}^c_m ) \, ,
\end{equation}
for each critical (or extremal) density matrix $\hat{\rho}^c_m$, with $m =1, 2, \ldots d\,!$. For the pure case the expectation values yield the energy spectrum of the system and the extremal density matrices are orthogonal projectors.

\section{Degenerate case of $\hat{H} $} \label{Deg}

In this case the rank of the matrix $ \boldsymbol{M} $ satisfies that $ r < d(d-1)$, whose value is associated to the orbits of the Hamiltonian (see Table~\ref{tab}).  As the critical Bloch vector $\boldsymbol{\lambda}^c$ has $ n$ free variables with $n =d^2-1-r$ one then has that $n> d-1$. Thus, if $N$ denotes the number of variables appearing in the expression for the mean value of the Hamiltonian in the state $\hat{\rho}^c$, $ \langle \hat{H} \rangle^c ={\rm Tr}(\hat{H} \, \hat{\rho}^c)$, there are two cases to consider: 
\begin{itemize}
\item[i)] When $ 1 \leq N \leq d-1 $, one has to select $n-N$ components of the density matrix Bloch vector to have $d-1$ free variables and then solve the polynomial system of equations~(\ref{s2:eq3}).
\item[ii)] When $ d-1 < N \leq n $, one has only to pick up $d-1$ components of the density matrix Bloch vector from the set of $N$ elements,  and again to solve the mentioned polynomial equation.
\end{itemize}

In both cases $d-1$ free variables will be determined by the polynomial system~(\ref{s2:eq3}). The remaining $ n - (d-1) $ components can be taken equal to zero because they do not affect the commutator of $\hat{H}$ with $\hat{\rho}^c$. Specifically, if we are interested in the eigensystem, i.e., all the set of $\{ c_k=0 \}$, one can apply the following method recursively:
\begin{itemize}
\item[$\boldsymbol{1})$] Make zero the $n-(d-1)$ components, to solve the polynomial system~(\ref{s2:eq3}), whose solution give at least two extremal density matrices. 
\item[$\boldsymbol{2})$] Take the trace of the first set of solutions $\hat{\rho}^c_k$ with the commuting general density matrix $\hat{\rho}^c$, this yields a system of algebraic equations by asking the orthogonality conditions, i.e., ${\rm Tr} (\hat{\rho}^c \,  \hat{\rho}^c_k  ) =  0$, where $k$ is a label counting the number of solutions.
\item[$\boldsymbol{3})$] Substitute the solutions of the linear system, to express the new critical density matrix in terms of the free variables, where $d-1$ are fixed by means of the positivity conditions (all $\{ c_k=0 \}$) and the rest of the components can be taken equal to zero.
\item[$\boldsymbol{4})$] Return to step 1 and repeat the procedure again, until one gets $d$ orthogonal projectors.
\end{itemize}

Of course, one has in this case several solutions related with the degeneracy of the Hamiltonian in similar form as in the standard diagonalization procedure of a finite Hamiltonian matrix. Although this may seem arbitrary, ultimately it is related to the codimension conditions~\cite{keller,caspers}. This topic will be addressed in a future contribution.

Similarly to~(\ref{s3:eq19}), in all cases, the energy spectrum is given by
\begin{equation} \label{s3:eq20}
\langle \hat{H} \rangle_m^c \equiv {\rm Tr}( \hat{H}\,  \hat{\rho}^c_m ) \, ,
\end{equation}
for each critical density matrix $\hat{\rho}^c_m$, with $m =1, 2, \ldots d$.

\section{Examples for $d=2, 3$, and $4$}\label{examples}

\subsection{Case d=$2$.}

For $d=2$, one has that the generators $\{\hat{\lambda}_k\}$ can be realized in terms of the Pauli matrices, i.e., $\hat{\lambda}_1 = \hat{\sigma}_1$, $\hat{\lambda}_2 = \hat{\sigma}_2$ and $\hat{\lambda}_3 = \hat{\sigma}_3$. Therefore the density and Hamiltonian matrices can be written in terms of the Bloch vectors $\boldsymbol{\lambda} = (\lambda_1, \lambda_2, \lambda_3)$ and $\boldsymbol{h} = (h_1, h_2, h_3)$,  
\begin{eqnarray} 
 \hat{\rho}&=&\frac{1}{2} \left(
\begin{array}{cc}
	1+\lambda_3 & \lambda_1-i\,\lambda_2 \\
	\lambda_1+i\,\lambda_2 & 1-\lambda_3
\end{array} \right) \, , \quad
\hat{H}=\frac{1}{2} \left(
\begin{array}{cc}
	h_0+h_3 & h_1-i\,h_2 \\
	h_1+i\,h_2 & h_0-h_3
\end{array} \right) \, ,
 \label{s5:eq1}
\end{eqnarray}
where $\boldsymbol{\lambda}$ is also called the polarization vector. 

Thus by substituting the expressions~(\ref{s5:eq1})  into~(\ref{s3:eq13}) we obtain the condition that the Bloch vectors of $H$ and the density matrix are parallel,
\begin{eqnarray}
(\boldsymbol{\lambda} \times \boldsymbol{h})_q = 0 \, , \nonumber
\end{eqnarray}
Its solution gives the critical Bloch vector
\begin{eqnarray} \label{s5:eq5}
\boldsymbol{\lambda}^c = \left( \frac{h_1}{h_3} \lambda_3, \, \frac{h_2}{h_3} \lambda_3, \, \lambda_3 \right) \, .
\end{eqnarray}
with a free variable $\lambda_3 $, according with the dimensions of the orbits of the Hamiltonian~(see Table~\ref{tab}). 

From expressions~(\ref{s2:eq3}), one has a single positivity condition
\begin{equation*}
c_2 = \frac{1}{4} (1 - (\lambda_1^c)^2 - (\lambda_2^c)^2 - \lambda_3^2) \, ,
\end{equation*}
by substituting~(\ref{s5:eq5}) into the above equation, we obtain
\begin{equation}\label{s5:eq7}
\lambda_3 = \pm \frac{\delta \, h_3}{h} \, ,
\end{equation}
where we define $h =\sqrt{ h_1^2+h_2^2+h_3^2}$ and $\delta=\sqrt{1-4c_2}$. By means of ~(\ref{s5:eq5}) and (\ref{s5:eq7}), we find the critical density matrices
\begin{equation}\label{s5:eq8} 
\hat{\rho}_{\pm}^c =  \frac{1}{2} \left(
\begin{array}{cc}
 1 \pm \frac{\delta \, h_3}{h}  & \pm \frac{\delta \, (h_1 -  i h_2 )}{ h} \\ 
 \pm \frac{\delta \, (h_1 + i h_2 )}{ h} &  1\mp \frac{\delta \,  h_3}{h} \\
\end{array} 
\right)  \, ,
\end{equation}
which correspond exactly to the solutions given in~\cite{figueroa}. Note that $\hat{\rho}^c_+ \hat{\rho}^c_-= c_2 \, I_2$. Substituting them into~(\ref{s3:eq19}) we get
\begin{eqnarray}
\langle \hat{H} \rangle_{\pm}^c &=& \frac{1}{2} \left(h_0 \pm \delta h \right) \, .
\label{s5:eq9} 
\end{eqnarray}	
We can distinguish two types of solutions: 
\begin{itemize}
\item Pure case (when $c_2=0$): One has $\delta=1$, the eigenvalues $\epsilon_{\pm} =\frac{1}{2} \left(h_0 \pm h \right)$ of the Hamiltonian, and from~(\ref{s5:eq8}), with $\delta=1$, the corresponding orthogonal projectors.
\item Mixed case (when $0< c_2 \leq1/4$): The extremal density matrices for the expectation value of the Hamiltonian are given in terms of the convex sum
\begin{equation}
\hat{\rho}^{\rm mixed}_\pm = \frac{1}{2} (1+\delta) \hat{\rho}^{\rm pure}_\pm + \frac{1}{2} (1-\delta) \hat{\rho}^{\rm pure}_\mp \, ,
\end{equation}
where $p_{\pm}=\frac{1}{2} (1+\delta)$ indicates the probability of finding the system with eigenvalue $\epsilon_+$ while $p_{\mp}=\frac{1}{2} (1-\delta)$ the corresponding probability of finding an energy $\epsilon_-$.
\end{itemize}

\subsection{Case d=$3$.} \label{sec3}

For the qutrit case, the generators $\hat{\lambda}_k$, with $k=1,2,\ldots 8 $ can be realized in terms of the Gell-Mann matrices~\cite{hioe}. Thus, an arbitrary density matrix is given by
\begin{eqnarray}
\hat{\rho} = \frac{1}{2} \left(
\begin{array}{ccc}
 \lambda_7+\frac{\lambda_8}{\sqrt{3}}+\frac{2}{3} & \lambda_1 - i \lambda_4 & \lambda_2 -i \lambda_5 \\
\lambda_1 + i \lambda_4 &  \frac{\lambda_8}{\sqrt{3}}-\lambda_7+\frac{2}{3} & \lambda_3 -i \lambda_6 \\
 \lambda_2 + i \lambda_5 & \lambda_3+ i \lambda_6  & \frac{2}{3}- \frac{2}{\sqrt{3}}\lambda_8 \\
\end{array}
\right) \, , \label{s5:eq12}
\end{eqnarray}
while the matrix~(\ref{s3:eq16}) takes the form 
\begin{eqnarray}
\boldsymbol{M} = \frac{1}{2} \left(
\begin{array}{cccccccc}
 0 & h_6 & h_5 & 2 h_7 & -h_3 & -h_2 & -2 h_4 & 0 \\
 -h_6 & 0 & h_4 & -h_3 & h_{78} & h_1 & -h_5 & -\sqrt{3} h_5 \\
 -h_5 & -h_4 & 0 & h_2 & h_1 &h_{87} & h_6 & -\sqrt{3} h_6 \\
 -2 h_7 & h_3 & -h_2 & 0 & h_6 & -h_5 & 2 h_1 & 0 \\
 h_3 & -h_{78} & -h_1 & -h_6 & 0 & h_4 & h_2 & \sqrt{3}h_2 \\
 h_2 & -h_1 & -h_{87} & h_5 & -h_4 & 0 & -h_3 & \sqrt{3} h_3 \\
 2 h_4 & h_5 & -h_6 & -2 h_1 & -h_2 & h_3 & 0 & 0 \\
 0 & \sqrt{3} h_5 & \sqrt{3} h_6 & 0 & -\sqrt{3} h_2 & -\sqrt{3} h_3 & 0 & 0 \\
\end{array}
\right) \, , \label{s5:eq13} 
\end{eqnarray}
which is a real skew symmetric matrix, and to simplify the matrix notation we define $h_{78}=h_7 + \sqrt{3} \, h_8$ and $h_{87} =  \sqrt{3} \, h_8-h_7$.

We consider the following Hamiltonian matrix
\begin{eqnarray}
\hat{H} = 
\left(
\begin{array}{ccc}
	                        b   &  \frac{c}{\sqrt{2}}  	& 0 \\
	      \frac{c}{\sqrt{2}}  & 0  			        & \frac{c}{\sqrt{2}}   \\
	                          0	& \frac{c}{\sqrt{2}}     &  b
\end{array} \right) \, , \label{s5:eq14}
\end{eqnarray}
where the parameters $b$ and $c$ are real parameters.  It represents a Hamiltonian written in terms of the angular momentum $\hat{H}= b \, \hat{J}^2_z + c\, \hat{J}_x$ with $j=1$. This Hamiltonian has been used to describe
a two mode Bose-Einstein condensate where the parameter $b$ represents the atom-atom interaction, and $c$ is related with the tunnelling parameter or a symmetric system of two interacting qubits~\cite{citlali}. The Bloch vector for the Hamiltonian is given by 
\begin{eqnarray}
 \boldsymbol{h} = (\sqrt{2} \, c, \, 0, \, \sqrt{2} \, c,\, 0, \, 0, \, 0, \, b, \, -\sqrt{3} \, b)  \, . \nonumber
\end{eqnarray}
Substituting the components of $\boldsymbol{h}$ into~(\ref{s5:eq13}), one finds that the rank of $\boldsymbol{M}$ is $6$, implying that the Hamiltonian is non degenerate. Solving the system of equations~(\ref{s3:eq15}) with the Gauss-Jordan elimination method, one obtains the extremal Bloch vector for the density matrix 
\begin{eqnarray}
\boldsymbol{\lambda}&=&\Biggl( \frac{c \, \lambda_2}{\sqrt{2}\, b} -\frac{\sqrt{6} \, c \, \lambda_8}{b}, \lambda_2,\frac{c \, \lambda_2}{\sqrt{2}\, b} -\frac{\sqrt{6} \, c \, \lambda_8}{b}, \, 0,\, 0,\,  
 0,-\sqrt{3} \, \lambda_8, \lambda_8 \Biggr) \, . \nonumber
\end{eqnarray}
Thus the associated critical density matrix is found by replacing the components of $\boldsymbol{\lambda}^c$ into~(\ref{s5:eq12}), which is denoted by $ \hat{\rho}^{c}$. For this case, to guarantee the positivity of the density matrix, one must consider
\begin{eqnarray}
c_2 = \frac{1}{2} \left( 1 - {\rm Tr} ( \hat{\rho}^{c})^2  \,  \right)  \, , \quad
c_3 = \det \hat{\rho}^{c} \, . \label{s5:eq23}
\end{eqnarray}

\begin{figure}
\begin{center}
(a) \includegraphics[width=0.3 \textwidth]{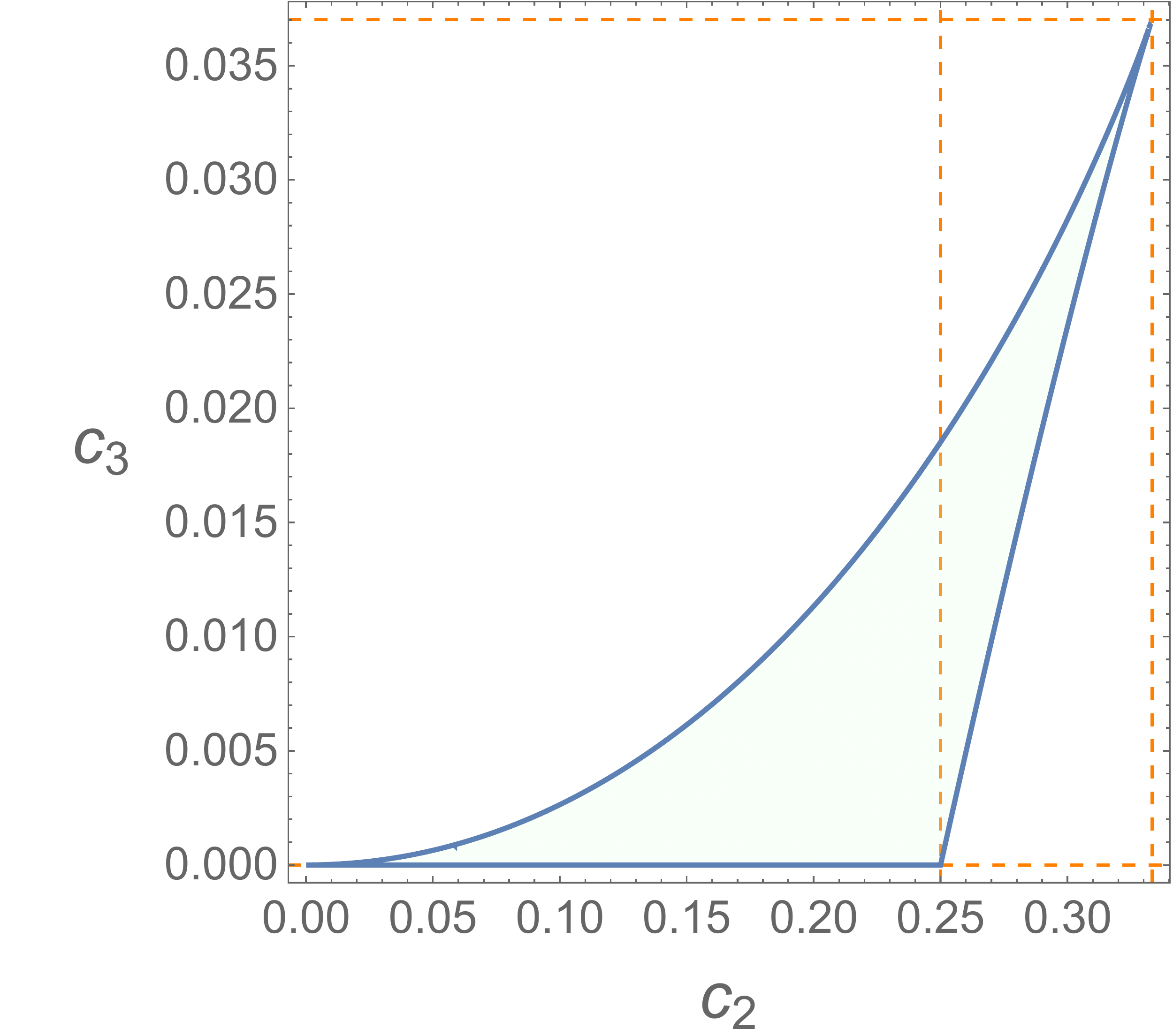} \qquad
(b) \includegraphics[width=0.45 \textwidth]{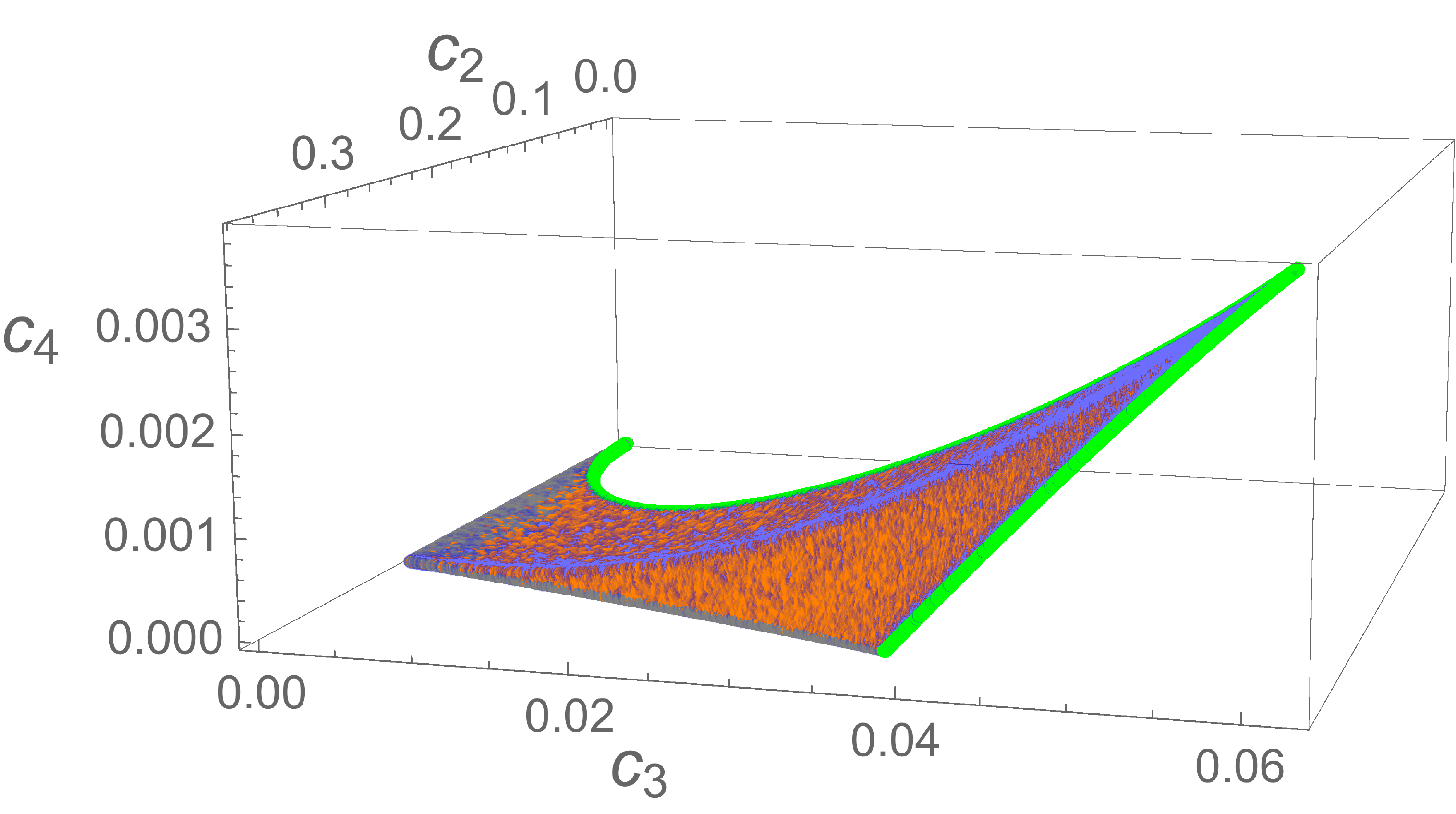}
\end{center}
\caption{ Regions of $c_2$, $c_3$, and $c_4$ where the positivity conditions of density matrix are satisfied. (a) For the case $d=3$, with the horizontal and vertical dashed lines taking the values $c_3=\{0,1/27\}$ and $c_2=\{1/4,1/3\}$, respectively.  (b) For the case $d=4$ one gets a solid figure. The pure case is associated to $(c_2,c_3,c_4)=(0,0,0)$ while the maximal mixed state correspond to $(c_2,c_3,c_4)=(3/8,1/16,1/256)$.} 
\label{region}
\end{figure}

Newly one has two types of solutions:
\begin{itemize}
\item Pure case (taking $c_2=c_3=0$): Solving the system of equations $c_2=0$ and $c_3=0$, one arrives to $3$ different solutions for $\lambda_2$ and $\lambda_8$, denoted by
\begin{eqnarray}
(\lambda_2, \lambda_8)_0 &=& \Bigl(-1,  -\frac{1}{2 \sqrt{3}}\Bigr) \, ,  \\
(\lambda_2, \lambda_8)_\pm &=& \frac{1}{2}\Bigl( 1 \pm \frac{b}{\sqrt{b^2 + 4 \, c^2}}, \frac{\sqrt{3}} {6}  \mp\frac{\sqrt{3}\, b}{2 \, \sqrt{b^2 + 4 \, c^2}} \Bigr) \, ,
 \end{eqnarray}
 which yield three independent Bloch vectors of the density matrix, namely
 \begin{eqnarray}
\boldsymbol{\lambda}_0 &=& \Bigl( 0,-1,0,0,0,0,\frac{1}{2},-\frac{1}{2 \, \sqrt{3}} \Bigr) \, , \\
\boldsymbol{\lambda}_\pm &=& \Bigl( \pm \frac{\sqrt 2 \, c}{\sqrt{b^2 + 4 \, c^2}}, \frac{1}{2}  \pm \frac{b}{2 \,\sqrt{b^2 + 4 \,c^2}}, \pm \frac{\sqrt 2 \, c}{\sqrt{b^2 + 4\, c^2}},0,0,0, \nonumber \\
&-&\frac{1}{4} \pm \frac{3\, b}{4 \,\sqrt{b^2 + 4 \,c^2}}, \frac{\sqrt{3}}{12} \mp \frac{\sqrt{3} \, b}{4\,\sqrt{b^2 + 4 \, c^2}} \Bigr) \, ,
\end{eqnarray}
whose norm is equal to $4/3$ and the scalar products between them are equal to $-2/3$. Thus it is straightforward to check that the corresponding extremal density matrices are orthogonal projectors associated to the energy eigenvalues of the Hamiltonian $\epsilon_0=b$, $\epsilon_\pm= 1/2( b \pm \sqrt{b^2 + 4 \, c^2})$.
\item Mixed case: For any other values for $c_2$ and $ c_3 $ in the region shown in~Fig.~\ref{region}(a), one can solve the polynomial system given by (\ref{s5:eq23}). As an example we take $c_2=29/100$ and $c_3=1/50$. There are $6$ different solutions for $\lambda_2$ and $\lambda_8$ which give rise to $6$ Bloch vectors,
\begin{eqnarray}
\boldsymbol{\lambda^{(1)}}_\pm &=& \Bigl( \pm \frac{\sqrt 2 \, c}{10\sqrt{b^2 + 4 \, c^2}}, \frac{7}{20}  \pm \frac{b}{20 \,\sqrt{b^2 + 4 \,c^2}}, \pm \frac{\sqrt 2 \, c}{ 10 \, \sqrt{b^2 + 4\, c^2}},0,0,0, \nonumber \\
&-&\frac{7}{40} \pm \frac{3\, b}{40 \,\sqrt{b^2 + 4 \,c^2}}, \frac{\sqrt{3}}{120} \mp \frac{\sqrt{3} \, b}{40 \, \sqrt{b^2 + 4 \, c^2}} \Bigr)  \, , \\
\boldsymbol{\lambda^{(2)}}_\pm &=& \Bigl( \pm \frac{2\, \sqrt 2 \, c}{5 \, \sqrt{b^2 + 4 \, c^2}}, -\frac{1}{10}  \pm \frac{b}{5 \,\sqrt{b^2 + 4 \,c^2}}, \pm \frac{2\, \sqrt 2 \, c}{5 \, \sqrt{b^2 + 4\, c^2}},0,0,0, \nonumber \\
&+&\frac{1}{20} \pm \frac{3\, b}{10 \,\sqrt{b^2 + 4 \,c^2}}, -\frac{\sqrt{3}}{60} \mp \frac{\sqrt{3} \, b}{10 \,\sqrt{b^2 + 4 \, c^2}} \Bigr) \, , \\
\boldsymbol{\lambda^{(3)}}_\pm &=& \Bigl( \pm \frac{3\, \sqrt 2 \, c}{10 \, \sqrt{b^2 + 4 \, c^2}}, -\frac{1}{4}  \pm \frac{3\, b}{20 \,\sqrt{b^2 + 4 \,c^2}}, \pm \frac{3\, \sqrt 2 \, c}{10 \, \sqrt{b^2 + 4\, c^2}},0,0,0, \nonumber \\
&+&\frac{1}{8} \pm \frac{9\, b}{40 \,\sqrt{b^2 + 4 \,c^2}}, -\frac{5 \, \sqrt{3}}{120} \mp \frac{9\, \sqrt{3} \, b}{120 \,\sqrt{b^2 + 4 \, c^2}} \Bigr) \, ,
\end{eqnarray}
whose corresponding extremal expectation values of the Hamiltonian are given by
\begin{eqnarray}
\langle \hat{H} \rangle^{(1)}_\pm &=& \frac{11\, b}{20}  \pm \frac{\sqrt{b^2 + 4 \, c^2}}{20} \, , \quad \langle \hat{H} \rangle^{(2)}_\pm= \frac{7\, b }{10} \pm \frac{\sqrt{b^2 + 4 \, c^2}}{5} \, , \nonumber \\
&& \langle \hat{H} \rangle^{(3)}_\pm = \frac{3\, b }{4}  \pm \frac{3\, \sqrt{b^2 + 4 \, c^2}}{20} \, .
\label{mixto3}
\end{eqnarray}
\end{itemize}

We find the expansion of the extremal density matrices for the mixed case in terms of the pure case described before,
\begin{eqnarray}
\hat{\rho}^{(1)}_+ &=& \frac{1}{10} \,\hat{\rho}_0 + \frac{1}{2} \, \hat{\rho}_+ + \frac{2}{5} \, \hat{\rho}_- \, , \qquad 
\hat{\rho}^{(1)}_- = \frac{1}{10} \, \hat{\rho}_0 + \frac{2}{5} \, \hat{\rho}_+ + \frac{1}{2} \, \hat{\rho}_- \, , 
\nonumber \\
\hat{\rho}^{(2)}_+ &=& \frac{2}{5} \,\hat{\rho}_0 + \frac{1}{2} \,\hat{\rho}_+ + \frac{1}{10} \, \hat{\rho}_- \, , \qquad 
\hat{\rho}^{(2)}_- = \frac{2}{5} \, \hat{\rho}_0 + \frac{1}{10} \, \hat{\rho}_+ + \frac{1}{2} \, \hat{\rho}_- \, , \nonumber \\
\hat{\rho}^{(3)}_+ &=& \frac{1}{2} \, \hat{\rho}_0 + \frac{2}{5} \,\hat{\rho}_+ + \frac{1}{10} \, \hat{\rho}_- \, , \qquad 
\hat{\rho}^{(3)}_- = \frac{1}{2} \, \hat{\rho}_0 + \frac{1}{10} \,\hat{\rho}_+ + \frac{2}{5} \, \hat{\rho}_- \, . 
\end{eqnarray}
Note that the expressions~(\ref{mixto3}) can be checked by calculating the expectation value of the Hamiltonian with the expansions given in the last expression.

\subsubsection{Degenerate case.}\label{sec3d}

Now we consider the Hamiltonian matrix given by
\begin{eqnarray}
\hat{H} = 
\left(
\begin{array}{ccc}
	2   				& - 1 +   i  	& - 1 - \frac{i}{3}   \\
	- 1 - i  	& \frac{13}{3}     						&1 + i \, 2   \\
	- 1 + \frac{i}{3}   					& 1 - i \, 2  		& 3
\end{array} \right) \, . \label{s5:eq14}
\end{eqnarray}
In this case the Bloch vector characterising the Hamiltonian is given by
\begin{equation}
\boldsymbol{h}= \Biggl(-2,\, -2,\, 2,\, -2,\, \frac{2}{3},\, -4,\, -\frac{7}{3},\, \frac{\sqrt{3}}{9}\Biggr)
\end{equation}
with $h_0 = \frac{28}{3}$. Replacing this values into the matrix~(\ref{s5:eq13}), the rank of $\boldsymbol{M} $ is $r=4$, which according to Table~\ref{tab} the Hamiltonian exhibits a double degeneracy. Thus, if $\alpha > \beta $ the diagonal representation is $ {\rm diag} (\alpha ,  \beta , \beta )$, or in opposite way, if $ \beta > \alpha $ then ${\rm diag} ( \beta ,\beta ,\alpha ) $.

By applying the Gauss-Jordan method to~(\ref{s3:eq15}), it yields
{\small
\begin{eqnarray}
\lambda_1^c &=& \frac{1}{14} \left(6 \,\lambda_5 +8 \, \lambda_6 + \lambda_7+7 \, \sqrt{3} \, \lambda_8 \right) , \, \nonumber \\
\lambda_2^c &=& \frac{1}{7} \left(3 \, \lambda_5 -3 \, \lambda_6 +11 \, \lambda_7 -7 \, \sqrt{3} \, \lambda_8 \right)  ,\nonumber \\
\lambda_3^c &=& \frac{1}{42} \left(24 \, \lambda_5 -24 \, \lambda_6 +11 \, \lambda_7 -7 \, \sqrt{3} \, \lambda_8 \right) , \, \nonumber \\
\lambda_4^c &=& \frac{1}{42} \left(-30 \, \lambda_5 +30 \, \lambda_6 -19 \, \lambda_7 +35 \, \sqrt{3} \,\lambda_8 \right)  \, .  \nonumber
\end{eqnarray} }
Hence, the corresponding critical Bloch vector~(\ref{s3:eq17}) is given by
\begin{eqnarray}
\boldsymbol{\lambda}^c = (\lambda_1^c,\, \lambda_2^c,\, \lambda_3^c,\,  \lambda_4^c ,\,  \lambda_5 ,\,  \lambda_6 ,\,     \lambda_7 ,\, \lambda_8  ) \, . \nonumber
\end{eqnarray}
with $4$ free parameters and its associated critical density matrix is denoted as $\hat{\rho}^{c}$.

Now, in order to obtain the eigensystem of $\hat{H}$, we are going to use the procedure established before for the degenerated case:  
\begin{itemize}
\item Thus we select the components $\lambda_5=\lambda_6=0$, solve the polynomial condition~(\ref{s5:eq23}) with $c_2=c_3=0$, and we get the following Bloch vectors of the density matrix
\begin{eqnarray}
\boldsymbol{\lambda}^{(1)} &=& \frac{1}{11} \biggl( 6, \, 0,\, 0,\, 6,\, 0,\, 0,\, 7, \,11/\sqrt{3} \biggr) \, , \ 
\boldsymbol{\lambda}^{(2)} = \frac{1}{46} \biggl( 12, \, 36,\, 6,\, 0,\, 0,\, 0,\, 35, \,19/\sqrt{3} \biggr) \, . \qquad
\end{eqnarray}
These Bloch vectors yield two density matrices $\hat{\rho}^{(1)}$, and $\hat{\rho}^{(2)}$ which are not independent, both by taking the trace with the Hamiltonian give an energy eigenvalue $\epsilon=4/3$.  
\item We establish the algebraic system of equations, 
\begin{equation}
\left\{ {\rm Tr}{\biggl(\hat{\rho}^{(1)}\, \hat{\rho}^c\biggr)}, {\rm Tr}{\biggl(\hat{\rho}^{(2)}\,\hat{\rho}^c\biggr)} \right\} =0 \, ,
\end{equation}
whose solution together with the positivity condition gives another Bloch vector
\begin{equation}
\boldsymbol{\lambda}^{(3)} = \frac{1}{16} \biggl( -6, \, -6,\, 6,\, -6,\, 2,\, -12,\, -7, \,1/\sqrt{3} \biggr) \, .
\end{equation}
Therefore we have obtained another extremal density matrix orthogonal to $\hat{\rho}^{(1)}$, and $\hat{\rho}^{(2)}$ and the expectation value of the Hamiltonian yields the eigenvalue $\epsilon_2=20/3$. Until now we have obtained $2$ independent and orthogonal projectors, we chose $\hat{\rho}^{(1)}$, and $\hat{\rho}^{(3)}$.
\item We repeat the procedure by establishing the algebraic system of equations
\begin{equation}
\left\{ {\rm Tr}{\Bigl(\hat{\rho}^{(1)} \, \hat{\rho}^c\Bigr)}, {\rm Tr}{\Bigl(\hat{\rho}^{(3)}\,\hat{\rho}^c\Bigr)} \right\} =0 \, ,
\end{equation}
whose solution give the Bloch vector
\begin{equation}
\boldsymbol{\lambda}^{(4)} = \Biggl( -\frac{15}{88} , \, \frac{3}{8} ,\,- \frac{3}{8} ,\, -\frac{15}{88} ,\,-\frac{1}{8} ,\, \frac{3}{4} ,\, -\frac{35}{176} , \,-\frac{17}{16\, \sqrt{3}}  \Biggr) \, .
\end{equation}
Thus one gets another orthogonal projector $\hat{\rho}^{(4)}$ and the expectation value of the Hamiltonian is $\epsilon_3=4/3$. 
\end{itemize}
We have obtained the complete eigensystem of the degenerated Hamiltonian. For the eigenvalue $\epsilon=4/3$, we indeed have a family of projectors yielding the same eigenvalue. This family is associated to the standard problem, when there is degeneracy, of the diagonalization of a Hamiltonian matrix, i.e., we can take any linear combination of the corresponding independent eigenstates.

\subsection{Case d=$4$.} \label{sec4}

For the states space of a quartit, the density matrix is given by
{\small
\begin{eqnarray}
\hat{\rho} = \frac{1}{2}
\left(
\begin{array}{cccc}
r_{11}	&\lambda_{1}-i\lambda_{7}	&\lambda_{2}-i\lambda_{8}		&\lambda_{3}-i\lambda_{9} \\
\lambda_{1}+i\lambda_{7}	&\frac{1}{6}(3-6\lambda_{13} + 2\sqrt{3}\lambda_{14} + \sqrt{6}\lambda_{15})	&\lambda_{4}-i\lambda_{10}	&\lambda_{5}-i\lambda_{11} \\
\lambda_{2}+i\lambda_{8}	&\lambda_{4}+i\lambda_{10}	&\frac{1}{6}(3-4\sqrt{3}\lambda_{14} + \sqrt{6}\lambda_{15})	&\lambda_{6}-i\lambda_{12} \\
\lambda_{3}+i\lambda_{9}	&\lambda_{5}+i\lambda_{11}		&\lambda_{6}+i\lambda_{12}	&\frac{1}{2}(1- \sqrt{6}\lambda_{15})
\end{array} \right)  , \qquad \label{sw:eq1}
\end{eqnarray} }
where we define $r_{11}=\frac{1}{6}(3+6\lambda_{13} + 2\sqrt{3}\lambda_{14} + \sqrt{6}\lambda_{15})$. 

We consider the Hamiltonian matrix
\begin{eqnarray}
\hat{H} = \left(
\begin{array}{cccc}
 a & \delta & b+a i & b+a i \\
 \delta & a & -b+ia & b-i a \\
 b-i a & -b-ia  & b & 0 \\
 b-i a & b+a i & 0 & b 
\end{array}
\right) \, , \label{sw:eq2}
\end{eqnarray}
with $a$, $b$ and $\delta$ as real parameters.

In the basis of the generalized Gell-Mann matrices $\hat{\lambda}_k$, with $k=1,2,\ldots 15$, the parameters Bloch vector for the Hamiltonian, $h_k = {\rm Tr} (\hat{H} \, \hat{\lambda}_k) $, is given by
\begin{eqnarray}
\boldsymbol{h} =2 \Biggl( \, \delta, \,  b, \,  b,\, - b,\,  b,\, 0, \, 0, \, - a, \, - a, \, - a, \, a, \, 0, \, 0,  \, \frac{\sqrt{3}}{3} (a-b),\, \sqrt{\frac{1}{6}} (a-b) \Biggr) \, . 
\label{sw:eq3} 
\end{eqnarray}
with $h_{0} =2(a+b)$. 

Therefore, its associated matrix~(\ref{s3:eq16}) is
\begin{eqnarray}
\boldsymbol{M} = 
\left(
\begin{array}{ccccccccccccccc}
 0 & -a & a & -a & -a & 0 & 0 & b & -b & -b & -b & 0 & 0 & 0 & 0 \\
 a & 0 & 0 & 0 & 0 & -a & b & r & 0 & \delta & 0 & -b & a & \eta & 0 \\
 -a & 0 & 0 & 0 & 0 & -a & -b & 0 & r & 0 & \delta & b & a & \frac{a}{\sqrt{3}} &  \alpha \\
 a & 0 & 0 & 0 & 0 & a & b & \delta & 0 & r & 0 & -b & -a & \eta & 0 \\
 a & 0 & 0 & 0 & 0 & -a & b & 0 & \delta & 0 & r & -b & a & -\frac{a}{\sqrt{3}} & - \alpha \\
 0 & a & a & -a & a & 0 & 0 & b & b & b & -b & 0 & 0 & 0 & 0 \\
 0 & -b & b & -b & -b & 0 & 0 & -a & a & a & a & 0 & 2 \delta & 0 & 0 \\
 -b & -r & 0 & -\delta & 0 & -b & a & 0 & 0 & 0 & 0 & a & b & \gamma & 0 \\
 b & 0 & -r & 0 & -\delta & -b & -a & 0 & 0 & 0 & 0 & -a & b & \frac{b}{\sqrt{3}} & \beta \\
 b & -\delta & 0 & -r & 0 & -b & -a & 0 & 0 & 0 & 0 & -a & b & -\gamma & 0 \\
 b & 0 & -\delta & 0 & -r & b & -a & 0 & 0 & 0 & 0 & -a & -b & \frac{b}{\sqrt{3}} & \beta \\
 0 & b & -b & b & b & 0 & 0 & -a & a & a & a & 0 & 0 & 0 & 0 \\
 0 & -a & -a & a & -a & 0 & -2\delta & -b & -b & -b & b & 0 & 0 & 0 & 0 \\
 0 & -\eta & -\frac{a}{\sqrt{3}} & -\eta & \frac{a}{\sqrt{3}} & 0 & 0 & -\gamma & -\frac{b}{\sqrt{3}} & \sqrt{3} b & -\frac{b}{\sqrt{3}} & 0 & 0 & 0 & 0 \\
 0 & 0 & - \alpha & 0 & \alpha & 0 & 0 & 0 & -\beta & 0 & - \beta & 0 & 0 & 0 & 0 \\
\end{array}
\right)  \, ,  \nonumber \label{sw:eq4}
\end{eqnarray} 

\noindent
where to simplify the matrix notation we have defined $r=a-b$, $\beta=\sqrt{8/3} \, b$, $\alpha=\sqrt{8/3} \, a$, $\gamma=\sqrt{3} \, b$, and $\eta=\sqrt{3} \, a$. We are going to consider two illustrative instances to exemplify the non degenerate and degenerate cases.

\subsubsection{Non degenerate case.} 

If $a=1$, $b=1/2$, and $\delta \neq 0$, thus the rank of $\boldsymbol{M} $ equals to $r=12$. In consequence, from Table~\ref{tab}, for these values the Hamiltonian~(\ref{sw:eq2}) is non degenerate. By applying the Gauss-Jordan method to the system~(\ref{s3:eq15}), one gets the Bloch vector of the density matrix, 
\begin{eqnarray}
\boldsymbol{\lambda}&=&\Bigl(\, \lambda^c_1,\, \frac{\lambda_{11}}{2},\, \lambda^c_3,\, -\frac{\lambda_{11}}{2},\, \lambda^c_3,\, 0,\, 0,\, -2\, \lambda^c_3, \, -\lambda_{11},
 -2\, \lambda^c_{3},\, \lambda_{11}, \lambda^c_{12},\, 0, \, \lambda_{14},\, \lambda_{15} \Bigr) \, ,
\end{eqnarray}
where 
\begin{eqnarray}
\lambda_{1}^c  &=& -\frac{1}{18} ( 9\, (1-2\, \delta) \lambda_{11} - 2 \sqrt{3} (\lambda_{14} + 8 \,\sqrt{2} \lambda_{15} )  \, , \nonumber \\
\lambda_{3}^c &=& \frac{  -3\, (1-2\, \delta) \lambda_{11} + 8 \, \sqrt{3} \lambda_{14} + 4 \, \sqrt{6} \lambda_{15} }{6 \, ( 1 + 2 \, \delta)} \, , \nonumber \\
\lambda_{12}^c &=& \frac{  -4\, \sqrt{3}}{9} \,(\lambda_{14} -  \, \sqrt{2} \lambda_{15} ) \, . 
\end{eqnarray} 

\begin{figure}
	\centering
	(a) \includegraphics[width=0.4\textwidth]{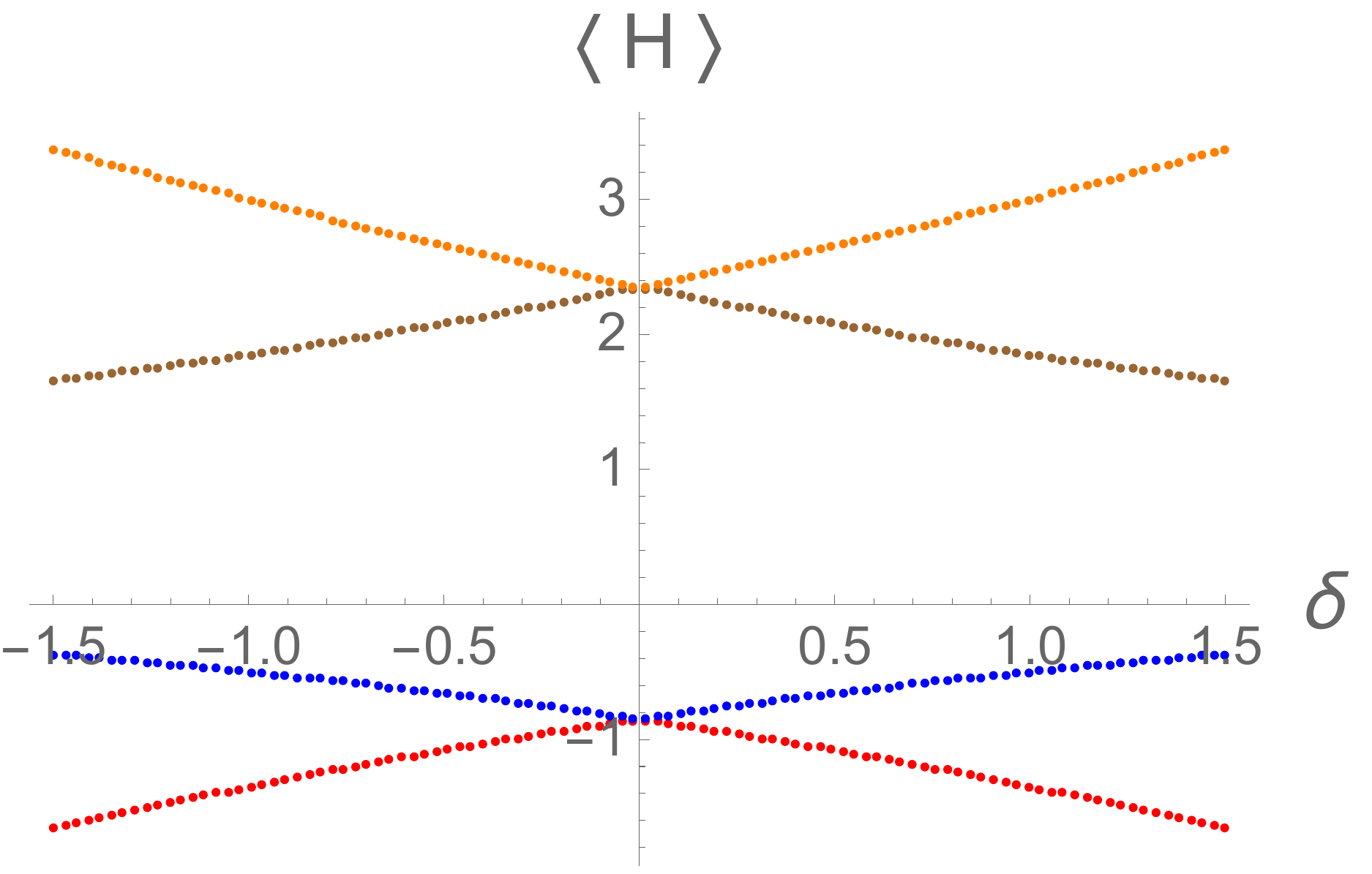}\qquad
	(b) \includegraphics[width=0.4\textwidth]{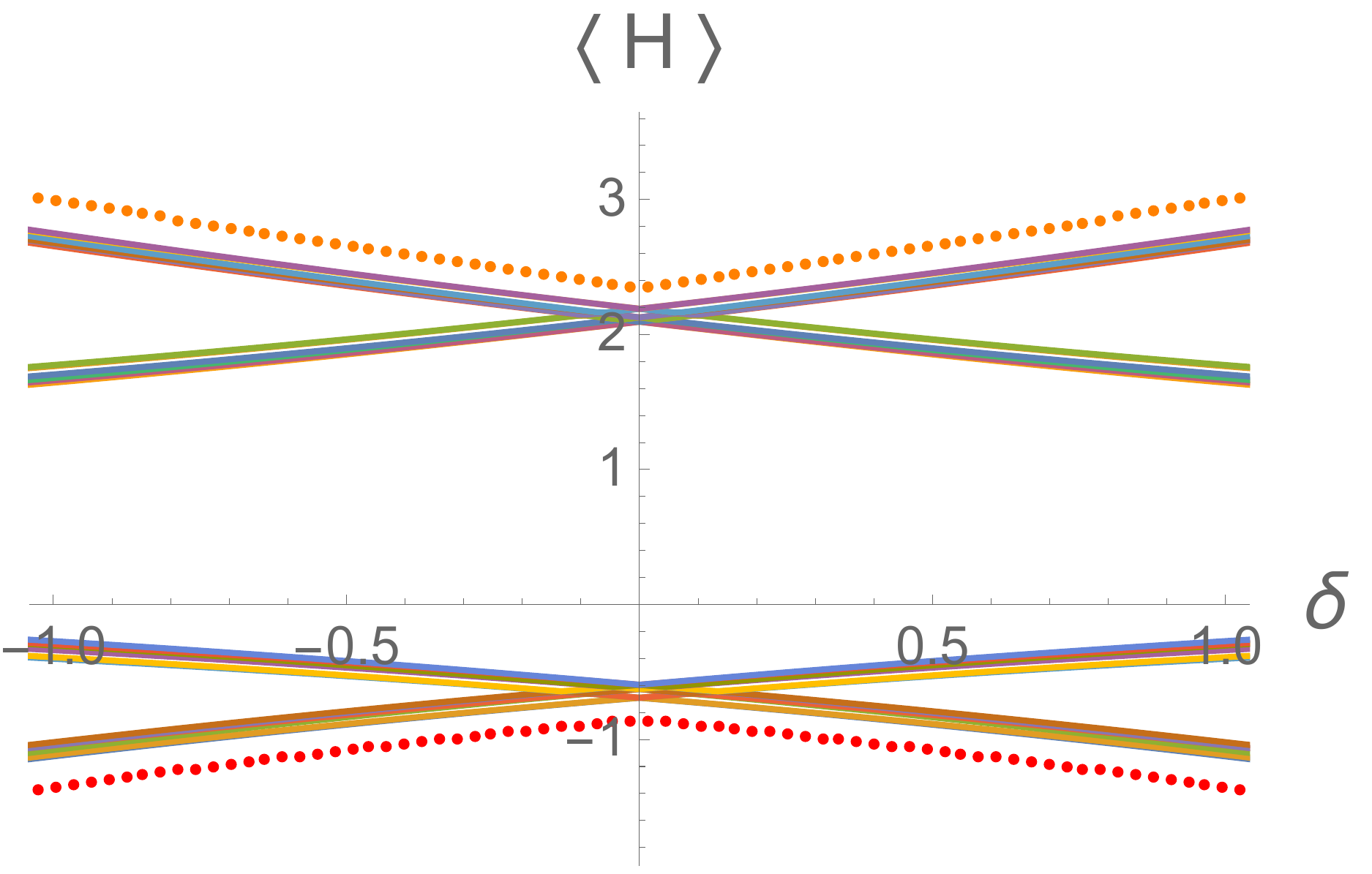}
\caption{ $ \langle \hat{H} \rangle^c $ as a function of $ \delta $. (a) Pure case with $c_2 = c_3 = c_4 = 0$; and (b) Mixed case with $c_2 = \frac{931}{10000}, \,  c_3 = \frac{141}{50000} ,\, c_4 = \frac{27}{1000000}$ are plotted with continuous lines. The dotted lines represent the minimum and maximum eigenvalues of the pure case.}
	\label{mixto}
\end{figure}

The components $\{\lambda_{11},\, \lambda_{14} ,\, \lambda_{15} \}$ are free variables, which are determined by establishing the system of polynomial equations~(\ref{s3:eq7}), where the constants $ c_2 $, $ c_3 $ and $ c_4 $ must lie inside the allowed region exhibited in Fig.~\ref{region}(b).  We consider two cases: (i) the pure case when one has $c_2 = c_3 = c_4 = 0$, which has four independent solutions for the parameters $(\lambda^c_{11},\lambda^c_{14},\lambda^c_{15})$. The extremal density matrices are projectors defined by $\hat{\rho}^c_{1 \pm} $ and $\hat{\rho}^c_{2 \pm} $. They are functions of the parameter $\delta$ and the corresponding expectation values are plotted in Fig~\ref{mixto}(b). The levels are indicated by dotted lines, which indicates that for $\delta=0$ the Hamiltonian system is degenerated by pairs.
(ii) The mixed case is established by taking from the region exhibited in Fig.~\ref{region}(c) the values $c_2 = \frac{931}{10000}, \,  c_3 = \frac{141}{50000} ,\, c_4 = \frac{27}{1000000}$. One has 24 different extremal expectation values of the Hamiltonian, $6$ for each energy level of the pure case. The results are shown also in Fig.~\ref{mixto}(b) with continuous lines. Notice that the extremal expectation values are contained within the minimum and maximum eigenvalues of the Hamiltonian.

\subsubsection{Degenerate case.}

For $\delta=0 $ in the Hamiltonian~(\ref{sw:eq2}), the rank of $\boldsymbol{M} $ is $r=8$ implying, from Table~\ref{tab}, that $\hat{H} $ is doubly degenerate and its diagonal representation is of the form $ {\rm diag} (\alpha , \alpha , \beta, \beta )$.

By applying the Gauss-Jordan method to the system~(\ref{s3:eq15}), one gets the Bloch vector of the density matrix with seven free components $(\lambda_7,\, \lambda_{10},\, \lambda_{11},\, \lambda_{12},\, \lambda_{13}, \, \lambda_{14}, \, \lambda_{15})$; the others can be written as 
\begin{eqnarray}
\lambda_{1}^c  &=& \frac{3 \, b \, (\lambda_{10} + \lambda_{11} +2 \, \lambda_{12}-2 \lambda_7 )-a \,\left(3 \lambda_{10} +3 \,\lambda_{11} +2 \,\sqrt{3} \lambda_{14} -2 \,\sqrt{6} \lambda_{15} \right)}{6 \,a} \, , \nonumber \\
\lambda_{2}^c  &=& \frac{a^2 \,(\lambda_{12} +\lambda_{7})+a \, b \, \left( \lambda_{10} +2 \, \sqrt{3} \,\lambda_{14} \right)-b^2 \, (\lambda_{10} +\lambda_{12} -\lambda_{7})}{a\, (a-b)} \, , \nonumber  \\
\lambda_{3}^c  &=& \frac{-3 \, a^2 (\lambda_{12} +\lambda_{7})+a \, b \left(-3 \, \lambda_{11} +2 \, \sqrt{3} \lambda_{14} +4 \, \sqrt{6} \lambda_{15} \right)+3 \, b^2 (\lambda_{11} +\lambda_{12} -\lambda_{7} )}{3 \, a \, (a-b)} \, , \nonumber  \\
\lambda_{4}^c  &=& \frac{-3 \, a^2 \, (\lambda_{12} +\lambda_{7})+a \, b \, \left(-3 \lambda_{11} -2 \sqrt{3} \lambda_{14} +2 \, \sqrt{6} \, \lambda_{15} \right)+3 \,  b^2 \, (\lambda_{11} +\lambda_{12} -\lambda_{7} )}{3 \, a \,  (a-b)} \, , \label{sw:eq5} \\
\lambda_{5}^c  &=&  \frac{-3 \, a^2 (\lambda_{12} +\lambda_{7})+a \, b \, \left(-3 \lambda_{10}-2 \sqrt{3} \lambda_{14}+2 \, \sqrt{6} \lambda_{15} \right)+3 \, b^2 (\lambda_{10} +\lambda_{12}-\lambda_{7})}{3 \, a \, (a-b)} \, , \nonumber  \\
\lambda_{6}^c  &=& \lambda_{13} -\frac{a \, \left(3 \lambda_{10} -3\, \lambda_{11} +4 \,\sqrt{3} \,\lambda_{14} +2 \,\sqrt{6} \,\lambda_{15} \right)+3 \, b \, (\lambda_{11} -\lambda_{10})}{6 \, a} \, , \nonumber  \\
\lambda_{8}^c  &=& \lambda_{11}-\frac{2 \, a \, \left(2 \lambda_{14}+\sqrt{2} \, \lambda_{15}\right)}{\sqrt{3} (a-b)} \, , \nonumber  \\
\lambda_{9}^c  &=& -\frac{2 \, a \, \left(2 \, \lambda_{14}+\sqrt{2} \, \lambda_{15}\right)}{\sqrt{3} \, (a-b)}-\lambda_{10} \, . \nonumber 
\end{eqnarray} 
The mean value of $\hat{H}$ in the state $\hat{\rho}^{c}$ depends only on the components $\lambda_{14}$ and $\lambda_{15}$; the remaining free variables can be chosen from the mentioned free set.  By considering $\{\lambda_{7},\, \lambda_{10} ,\, \lambda_{11},\, \lambda_{12} \} $ equal to zero, the components $\{\lambda_{13},\, \lambda_{14} ,\, \lambda_{15} \}$ are obtained by solving the system of polynomial equations~(\ref{s3:eq7}) in terms of the constants $ \{ c_2 , c_3 , c_4 \}$. 

For the pure case, associated to $c_2 = c_3 = c_4 = 0$, the set of solutions for this polynomial systems are 
\begin{eqnarray}
\{\lambda_{13}, \, \lambda_{14}, \, \lambda_{15}  \}_{1\pm} = \pm\Bigg\{  \frac{(a-b) \pm P}{2 P}  , \, \frac{a-b}{\sqrt{3} P} , \, \frac{a-b}{\sqrt{6} P}  \Bigg\}  \, , \nonumber 
\end{eqnarray} 
where $ P \equiv \sqrt{9 a^2-2 a b+9 b^2}$. Therefore, their respective critical density matrices are
\begin{eqnarray}
\hat{\rho}^c_{1\pm} = \pm \frac{1}{24 P} \left(
\begin{array}{cccc}
 12 \left(a-b \pm P\right) & 0 & 24 (b+ia ) & 24 (b+ia ) \\
 0 & 0 & 0 & 0 \\
 24 (b-ia) & 0 & 6 \left(b-a \pm P \right) & 6 \left(b-a \pm P \right) \\
 24 (b-ia) & 0 & 6 \left(b-a \pm P\right) & 6 \left(b-a \pm P \right)	
\end{array}
\right) \, ,\nonumber
\end{eqnarray}
which correspond to orthogonal projectors of $\hat{H}$ related to its degenerate eigenvalues, given respectively by
\begin{eqnarray}
{\rm Tr} ( \hat{\rho}^c_{1-} \,\hat{H} ) = \frac{1}{2} \left( a+b - P \right) \, , \quad
{\rm Tr} ( \hat{\rho}^c_{1+} \,\hat{H} ) = \frac{1}{2} \left( a+b + P \right) \, . \nonumber 
\end{eqnarray}

In order to find the remaining projectors, one constructs the linear system of equations
\begin{eqnarray}
{\rm Tr} (\hat{\rho}^c \,  \hat{\rho}^c_{1-}  )  = 0 \, , \quad 
{\rm Tr} (\hat{\rho}^c \,  \hat{\rho}^c_{1+}  )  = 0 \, , \nonumber
\end{eqnarray}
where the $\hat{\rho}^{c}$ is written in terms of the general density matrix that commutes with the Hamiltonian.

By solving it for $\lambda_{11}$ and $\lambda_{14}$ in terms of $\lambda_{10} , \, \lambda_{13}$ and $\lambda_{15}$, one finds
\begin{eqnarray}
\lambda_{11} &=& \frac{a \, (\lambda_{10}-8 \lambda_{13}-4)-b \, \lambda_{10}}{a-b} ,
\, \nonumber \\
\lambda_{14} &=& \ -\frac{6 \, \lambda_{13}+\sqrt{6} \, \lambda_{15}+3}{2 \, \sqrt{3}} \, . \nonumber
\end{eqnarray}
Then by replacing this solution into $\hat{\rho}^c$, setting $\lambda_{7}$ and $\lambda_{12}$ equal to zero, and solving the polynomial system (\ref{s3:eq7}) for  $c_2 = c_3 = c_4 = 0$ in terms of $\lambda_{10} , \, \lambda_{13}, \, \lambda_{15}$, one has
\begin{eqnarray}
\{\lambda_{10}, \, \lambda_{13}, \, \lambda_{15}  \}_{2\pm} =\pm\frac{1}{P} \Bigg\{  - 2 \, a , \,  \frac{b-a\mp P}{2} , \, \frac{a-b}{\sqrt{6}}  \Bigg\}  \, , \nonumber
\end{eqnarray} 
which yield the one rank projectors 
\begin{eqnarray}
\hat{\rho}^c_{2\pm} = \pm \frac{1}{24 P} \left(
\begin{array}{cccc}
 0 & 0 & 0 & 0 \\
 0 & 12 \, (a-b \pm P) & 24 \, (-b+i a) & 24 \, ( -i \, a +b) \\
 0 & -24 \, (b+a \, i) & 6\, (b-a \pm P) & 6 \, (a-b \mp P) \\
 0 & 24 \, (i \, a+b) & 6 \, (a-b \mp P) & 6 \, (b-a \pm P) \\
\end{array}
\right) \, . \nonumber
\end{eqnarray}
The respective expectation values of the Hamiltonian are
\begin{eqnarray}
{\rm Tr} ( \hat{\rho}^c_{2-} \,\hat{H} ) = \frac{1}{2} \left( a+b - P \right) \, , \quad
{\rm Tr} ( \hat{\rho}^c_{2+} \,\hat{H} ) = \frac{1}{2} \left( a+b + P \right) \, . \nonumber
\end{eqnarray}

Finally, it is possible to corroborate that the set $ \{ \hat{\rho}^c_{1\pm}, \, \hat{\rho}^c_{2\pm} \} $ are a complete set of orthogonal one rank projectors because $ \hat{\rho}^c_{1+} + \hat{\rho}^c_{1-} + \hat{\rho}^c_{2+} + \hat{\rho}^c_{2-} = \hat{I}_{4}$.

\section{Summary and Conclusions}

The main contribution of our work is to give an algebraic procedure to find extremal density matrices for a given Hamiltonian. Our approach applies to both the degenerate and non degenerate cases of the Hamiltonian. The examples of the procedure are given for dimensions $d=2,\,3,\,4,$ and show that the Hamiltonian spectrum for the pure case is recovered. For the mixed case, we have verified that the extremal values of the expectation value of the Hamiltonian is a convex sum of the corresponding results for the pure case. We want to enhance that the method can be applied by replacing the Hamiltonian for any observable acting on a qudit space.

We established that an extremal density matrix commutes with the Hamiltonian operator and optimises its mean value. We demonstrated that at most $d-1$ variables are necessary to find extremal density matrices with appropriate positivity conditions, for the non-degenerated case of the finite matrix Hamiltonian. In the degenerate pure case, one has more free components of the extremal density matrix which can be selected by asking orthogonality between the projectors, which allow us to obtain the energy spectrum. 

Finally, in Appendix~\ref{B} following the method given in~\cite{gerdt}, we find also the compatible regions between the coefficients of the characteristic polynomial of the density matrix in terms of the positivity conditions of the Bezoutian matrix in order to provide a self-contained approach.

\section*{Acknowledgement}
This work was partially supported by CONACyT-M\'exico (under Project No.~238494) and DGAPA-UNAM (under Project No.~IN110114). The authors would like to thank Giuseppe Marmo, Margarita A. Man'ko and Vladimir I. Man'ko for their valuable comments and also to CONACyT-M\'exico for the Ph.D. scholarship to A.F.

\appendix

\section{Bezoutian matrix}\label{B} 

For $k=1,\ldots,d$, the elements $ t_k \equiv {\rm Tr} (\hat{\rho}^k ) $ form an integrity basis for all polynomial $U(d)$ invariants. In terms of them, it is defined the symmetric matrix called Bezoutian given in Eq.~(\ref{B:eq1}).

A polynomial with real coefficients has reals roots iff the Bezoutian matrix is positive definite~\cite{procesi}. Hence, the compatible region among the global invariants is obtained with the intersection of the positivity conditions of the density matrix from~(\ref{s2:eq1}) with the respective positivity conditions of the Bezoutian, mainly in its determinant $ \det \boldsymbol{B}_d  \geq 0 $~\cite{gerdt}. Besides, due to $ \det \boldsymbol{B}_d $ is equal to the discriminant of the characteristic polynomial of $ \hat{\rho} $, the degeneracy condition is obtained by the vanishing of $ \det \boldsymbol{B}_d $~\cite{deen,kurosch, bhattacharya}.  

On the other hand, the relation between the constants $ \{ c_p \} $ with $t_k $ is established by
\begin{equation}\label{B:eq2}
t_k = \sum_{p=1}^k (-1)^{p+1} c_p \, t_{k-p} \, , \qquad t_0 \equiv k  \, ,
\end{equation}
with $k=1,\ldots,d$. 

Next we establish the allowed regions of the $ \{ c_p \} $ and $t_k$ for the matrix Hamiltonians with dimensions $d=\,2,\,3,\,4$. Consequently, the Bezoutian matrix for $d=2$ is
\begin{equation*}
\boldsymbol{B}_2 = \left(
\begin{array}{cc}
	2      & 1 	\\
	1    & t_2  	
\end{array} \right) \, ,
\end{equation*}
where from formula~(\ref{B:eq2}), it is obtained $ t_2 = 1- 2 \, c_2 $. Then the condition $ \det \boldsymbol{B}_2  \geq 0 $ gives the positivity condition $c_2 \leq 1/4$, which corroborates the maximum value~(\ref{s2:eq1}).

For the case $d=3$, the positivity conditions~(\ref{s2:eq1}) of the density matrix are
\begin{eqnarray}
0 \leq c_2 = \frac{1}{2} \left( 1 - t_2 \right) \leq \frac{1}{3} \, , \label{B:eq3}\\
0 \leq c_3 = \frac{1}{6} \left( 1 - 3 t_2 + 2 t_3 \right) \leq \frac{1}{27} \, , \label{B:eq4}
\end{eqnarray}
while the Bezoutian matrix is
\begin{equation*}
\boldsymbol{B}_3 =  \left(
\begin{array}{ccc}
	3		& 1		& t_2 	\\
	1    	& t_2  	& t_3			\\
	t_2		& t_3 	& t_4  
\end{array} \right) \, .
\end{equation*} 

By applying the Cayley-Hamilton theorem and the formula~(\ref{B:eq2}) one obtains
\begin{eqnarray}
t_2  &=&  1 - 2 \, c_2 \, , \quad
t_3  =  1 - 3 c_2 + 3c_3 \, ,  \nonumber  \\
t_4  &=&  1-4c_2+2c_2^2 + 4c_3  \, . 
\label{B:eq5}
\end{eqnarray}
Similarly to the case of $d=2$, $ \det \boldsymbol{B}_3  \geq 0 $ is the only relevant positivity condition of $ \boldsymbol{B}_3 $. Expressed in terms of $ \{ c_2,\, c_3 \}$, it gives
\begin{eqnarray}
c_2^2 - 4 c_2^3 + 18 c_2 c_3 - c_3 (4 + 27 c_3) \geq 0 \, .\label{B:eq6}
\end{eqnarray}
Thus, the inequalities system formed by~(\ref{B:eq3}), (\ref{B:eq4}) and~(\ref{B:eq6}) produces the compatible region between $c_2$ and $c_3$. This is shown in Fig.~\ref{region}(a). The bottom line is associated with one eigenvalue zero (yielding the condition on $c_2$ for the case $d=2$) while the other curves imply two equal eigenvalues for the density matrix. The $(c_2,c_3)=(0,0)$ case is associated to density matrices of pure states and the highest is the maximal mixed state (all the eigenvalues are equal).

In the case of $d=4$, the positivity conditions of the density matrix are given by
\begin{eqnarray}
&& 0 \leq c_2 = \frac{1}{2} \left( 1 - t_2 \right) \leq \frac{3}{8} \, , \nonumber\\
&& 0 \leq c_3 = \frac{1}{6} \left( 1 - 3 t_2 + 2 t_3 \right) \leq \frac{1}{16} \, , \label{B:eq7} \\
&& 0 \leq c_4 = \frac{1}{24} \left( 1+ 3 t_2^2-6 t_2 +8 t_3 -6 t_4  \right) \leq \frac{1}{256} \, , \nonumber
\end{eqnarray}
and the respective Bezoutian matrix is
\begin{equation*}
\boldsymbol{B}_4 =  \left(
\begin{array}{cccc}
	4		& 1		& t_2 	& t_3	\\
	1    	& t_2  	& t_3	& t_4	\\
	t_2		& t_3 	& t_4	& t_5    \\
	 t_3	& t_4	& t_5	& t_6
\end{array} \right) \, ,
\end{equation*} 
with
\begin{eqnarray}
t_2 & = & 1 - 2 \, c_2 \, , \quad
t_3  =  1 - 3 c_2 + 3c_3 \, , \nonumber \\
t_4 & = & 2 (c_2-2) c_2 +4 c_3 +4 c_4 +1 \, , \label{B:eq8} \\
t_5 & = & 5 c_2 (c_2 -c_3 -1)+5 c_3 +5 c_4 +1 \, , \nonumber \\
t_6 & = & 9 c_2^2 - 2 c_2^3 -6 (2 c_3 + c_4 +1) c_2 +3 c_3 (c_3 +2)+6 c_4 +1 \, . \nonumber
\end{eqnarray}

In this case, $ \det \boldsymbol{B}_4  \geq 0 $ is the main condition; nevertheless, the remaining ones are crucial to avoid fake points in the compatible region for $ \{ c_2,\, c_3,\, c_4 \}$. All these conditions are

\bigskip

\begin{eqnarray}
&& 7 + 11 c_2^2 - 2 c_2^3 + c_3 (10 + 3 c_3) + 10 c_4 - 6 c_2 (2 + 2 c_3 + c_4) \geq 0 \, , \nonumber\\
\nonumber \\
&& c_2^4 + 30 c_3 - 4 c_2^5 + 42 c_4 + (c_3 (12 c_3 -17 ) - 17 c_4) (c_3 + c_4) + 2 c_2^3 (9 c_3 -8 - 10 c_4) + \nonumber\\
&& c_2^2 (33 + (4 - 19 c_3) c_3 + 18 c_4) - 2 c_2 (19 + 6 c_3 + 8 c_3^2 + 11 (1 + c_3) c_4 + 12 c_4^2) +9  \geq 0 \, , \nonumber \\ \nonumber \\
&& 8 c_3^3 - 2 c_3 (6 + c_4 (53 + 36 c_4))  - 27 c_3^4  + 2 c_3^2 (-37 + 9 c_4) +c_4 (6 - c_4 (77 + 64 c_4)) - \nonumber \\
&& 2 c_2^3 (8 + 2 (-9 + c_3) c_3 + 27 c_4) + c_2^2 (4 - 45 c_3^2 + 44 c_3 c_4 - 48 (c_4 - 1) c_4) -8 c_2^5 \label{B:eq9} \\  
&& 2 c_2 (c_3 (27 + c_3 ( 9 c_3 -8 )) - (3 + c_3 (43 + 45 c_3)) c_4 - 
23 c_4^2) + c_2^4 (2 - 8 c_4)  \geq 0 \nonumber \\ 
\nonumber \\
&& 18 c_2 (c_3 - 8 c_4) (c_3^2 - c_4) - c_3^3 (4 + 27 c_3) - 16 c_2^4 c_4 - 3 (9 + 64 c_3) c_4^2 + 4 c_2^3 (c_4-c_3^2) + \nonumber \\
&& 6 c_3^2 c_4  - 256 c_4^3  + c_2^2 (c_3^2 + 80 c_3 c_4 - 128 c_4^2) \geq 0 \, , \nonumber
\end{eqnarray}
where the last one is $ \det \boldsymbol{B}_4  \geq 0 $.

Hence, for the set $ \{ c_2,\, c_3,\, c_4 \}$, the region which satisfies the inequalities system formed by~(\ref{B:eq7}) and~(\ref{B:eq9}), is shown in Fig.~\ref{region}(b).  Notice that, by making zero $c_4$, we obtain the $d=3$ result, while by making zero two eigenvalues of the density matrix the line associated to the case $d=2$ is obtained ($c_3=c_4=0$). Inside the solid figure (orange color) one has the solution for $4$ eigenvalues of the density matrix different from zero whereas the surfaces are associated to $2$ degenerated eigenvalues (blue color). The curve for the case with three equal eigenvalues and the other different is also shown (green color).


\section*{References}

\begin{thebibliography}{20}
\bibitem{landau}
Landau L, 1927 Z. Phys. {\bf 45} 430. \\
ter Haar D, 1965 {\it Collected Papers of L.D. Landau} Pergamon Press.
\bibitem{dirac}
Dirac P.A.M. 1929 Proc. Cambridge Phil. Soc. {\bf 25} 62.
\bibitem{vonneumann}
von Neumann J, 1932 {\it Mathematische Grundlagen der Quantenmechanik} Berlin: Springer.\\
von Neumann J. 1955. {\it Mathematical Foundations of Quantum Mechanics} Princeton
University Press.
\bibitem{mahler}
Mahler G, Waberruss V A 1995 {\it Quantum Networks: Dynamics of Open Nanostructures.} Springer.
\bibitem{fano}
Fano U, 1953 Phys. Rev. {\bf 90} 577.\\
Fano U, 1957  Rev. Mod. Phys. {\bf 29} 74.
\bibitem{blum}
Blum K, 2012 {\it Density Matrix Theory and Applications} Springer.
\bibitem{newton}
Newton R, Young B L, 1968 Annals of Physics {\bf 49} 393.
\bibitem{park}
Park J L, Band W, 1971 Founds. of Physics {\bf 1} 211.
\bibitem{walser}
Walser R, Cirac J L, Zoller P, 1996. Phys. Rev. Lett. {\bf 77} 2658.
\bibitem{klose}
Klose G, Smith G, Jessen P S, 2001. Phys. Rev. Lett. {\bf 86} 4721.
\bibitem{amiet}
Amiet J P, Weigert S, 1999 J. Opt. B: Quantum Semiclass. Opt. {\bf 1} L5.
\bibitem{amiet2}
Amiet J P, Weigert S, 2000 J. Opt. B: Quantum Semiclass. Opt. {\bf 2} 118.
\bibitem{horn}
Horn R, Johnson C, 2013 {\it Matrix Analysis} Cambridge University Press.
\bibitem{chung}
Chung S U, Trueman T L, 1975. Phys. Rev. D {\bf 11} 633.
\bibitem{dodonov} 
Dodonov V V, Man'ko V I, 1997 Phys. Lett. A {\bf 229} 335.
\bibitem{olga}
Man'ko V I, Man'ko O V, 1997 J. Exp. Theor. Phys.  {\bf 85}, 430.
\bibitem{lopez1} 
Casta\~nos O, L\'opez-Pe\~na R, Man'ko M A, Man'ko V I, 2003, J. Phys. A: Mat. Gen. {\bf 36} 4677.
\bibitem{lopez2} 
Casta\~nos O, L\'opez-Pe\~na R, Man'ko M A, Man'ko V I, 2003, J. J. Opt. B: Semiclass. Opt. {\bf 5} 227.
\bibitem{bertlmann}
Bertlmann R., Krammer P, 2008 J. Phys. A: Math. Theor. {\bf 41} 235303.
\bibitem{hioe}
Hioe F T, Eberly J H, 1981. Phys. Rev. Lett. {\bf 47} 838.
\bibitem{aktha} 
Akhtarshenas S J, 2006 Optics and Spectroscopy {\bf 103} 411.
\bibitem{bruning} 
Br\"uning E, M\"akel\"a H, Messina A, Petruccione F, 2012 J. Mod. Opt. {\bf 59} 1.
\bibitem{jarls} 
Jarlskog C, 2006 J. Math. Phys. {\bf 47} 013507.
\bibitem{nielsen} 
Nielsen M A, Chuang I L, 2010 {\it Quantum Computation and Quantum Information} Cambridge University Press.
\bibitem{ritter}
Ritter W G. 2005. J. Math. Phys. {\bf 46} 082103.
\bibitem{giraud}
Giraud O, Braun D, Baguette D, Bastin T, Martin J, 2015. Phys. Rev. Lett. {\bf 114} 080401.
\bibitem{fujii} 
Fujii K, Funahashi K, Kobayashi T, 2006 Int. J. Geom. Methods Mod. Phys. {\bf 03} 269.
\bibitem{dita} 
Dita P, 2005 J. Phys A: Math. Gen. {\bf 38} 2657.
\bibitem{tilma} 
Tilma T, Sudarshan E C G, 2002 J. Phys A: Math. Gen. {\bf 35} 10467.
\bibitem{spengler} 
Spengler C, Huber M, Hiesmayr B, 2010 J. Phys. A: Math. Theor. {\bf 43} 385306.
\bibitem{petru} 
Br\"uning E, Chru\'sci\'nski D, Petruccione F., 2008 Open Syst. Inf. Dyn. {\bf 15} 397.
\bibitem{figueroa} 
Figueroa A., L\'opez J., Casta\~nos O., L\'opez-Pe\~na R, Man'ko M A, Man'ko V I,\\ 2015 J. Phys. A: Math. Theor. {\bf 48} 065301.
\bibitem{kimura} 
Kimura G, 2003 Phys. Lett. A {\bf 314} 339.
\bibitem{macfarlane}
Macfarlane A J, Sudbery A, Weisz P H 1968. Comm. Math. Phys. {\bf 11}, 77.
\bibitem{kimura2}
Kimura G, Kossakowski A, 2005. Open Sys. Information Dyn. {\bf 12} 207.
\bibitem{mallesh}
Mallesh K S, Mukunda N. 1997 Pramana J. Phys. {\bf 49} 371.
\bibitem{bengtsson}
Bengtsson I, Zyczkowski K, 2006 {\it Geometry of Quantum States. An introduction to Quantum Entanglement.} Cambridge University Press.
\bibitem{seroul}
Seroul R, 2000 {\it Programming for Mathematicians.} Springer.
\bibitem{byrd} 
Byrd M S, Khaneja N, 2003 Phys. Rev. A {\bf 68} 062322.
\bibitem{deen}
Deen S M, Kabir P K, 1971. Phys. Rev. D {\bf 4} 1662.
\bibitem{procesi}
Procesi C. 2007 {\it Lie Groups: An Approach through Invariants and Representations.} Springer.
\bibitem{gerdt} 
Gerdt V P, Khvedlidze A M, Palii Yu G, 2014 J. Math. Sci. {\bf 200} 682.
\bibitem{schwarz}
Procesi C, Schwarz G, 1985 Invent. Math. {\bf 81} 539.
\bibitem{lax}
Lax P D, 2007 {\it Linear Algebra and Its Applications.} Wiley-Interscience.
\bibitem{bhatia}
Bhatia R, 1997 {\it Matrix Analysis.} Springer.
\bibitem{gawron}
Gawron P, Puchala Z, Miszczak J A, Skowronek L, Zyczkowski K. 2010 J. Math. Phys. {\bf 51}, 102204.
\bibitem{waerden}
Van der Waerden B L, 1991 {\it Modern Algebra II. } Springer.
\bibitem{cox}
Cox D, Little J, O'Shea D, 2005 {\it Using Algebraic Geometry.} Springer.
\bibitem{gelfand}
Gelfand I M, Kapranov M M, Zelevinsky A V, 2008 {\it Discriminants Resultants and Multidimensional Determinants.} Birkh\"{a}user.
\bibitem{sturmfels}
Sturmfels B, 2002 {\it Solving Systems of Polynomial Equations.} Number 97, AMS Regional Conference Series 
\bibitem{helgason}
Helgason S, 1978 {\it Differential Geometry, Lie Groups, and Symmetric Spaces.} Academic Press Inc.
\bibitem{kus}
Kus M, Zyczkowski K, 2001 Phys. Rev. A {\bf 63} 032307.
\bibitem{linden1}
Linden N, Popescu S, Sudbery A, 1999 Phys. Rev. Lett. {\bf 83} 243.
\bibitem{ercolessi}
Ercolessi, E., Marmo, G., Morandi, G., 2001 Int. Jour. Mod. Phys. A {\bf 16} 5007.
\bibitem{schirmer}
Schirmer S G, Zhang T, Leahy J V, 2004 J. Phys. A: Math. Gen. {\bf 37} 1389.
\bibitem{gantmacher}
Gantmacher F R, 1959 {\it The Theory of Matrices Vol 1.} AMS Chelsea Publishing.
\bibitem{boya}
Boya L, Dixit K, 2008 Phys. Rev. A {\bf 78} 042108.
\bibitem{keller}
Keller J, 2008 Linear Algebra Appl. {\bf 429} 2209
\bibitem{caspers}
Caspers W J, 2008 J. of Phys.: Conference Series {\bf 104} 012032.
\bibitem{citlali} 
P\'erez-Campos C, Gonz\'alez-Alonso J R, Casta\~nos O, L\'opez-Pe\~na R, 2010 Ann. Phys. (N.Y.) {\bf 325} 325.
\bibitem{kurosch}
Kurosch A.G. 1984 {\it Higher Algebra.} MIR Publishers.
\bibitem{bhattacharya}
Bhattacharya M, Raman C, 2007 Phys. Rev. A {\bf 75} 033405.\\
Bhattacharya M, 2007 Am. J. Phys. {\bf 75} 942.
\end{thebibliography}
\end{document}